\documentclass[10pt, conference, compsocconf]{IEEEtran}

\usepackage{algorithm}
\usepackage{algorithmic}
\usepackage{amsfonts} 
\usepackage{color}
\usepackage[pdftex]{graphicx}
\usepackage{graphicx}
\usepackage[bookmarks=false]{hyperref}
\hypersetup{colorlinks=true,linkcolor=black,citecolor=black,filecolor=black,urlcolor=blue}
\usepackage{mathtools}
\usepackage{multirow}
\usepackage{stmaryrd} 
\usepackage{subcaption}
\usepackage{nicefrac}
\usepackage{amsmath}

\begin{document}

\title{Predicting computational reproducibility of data analysis pipelines \\in large population studies using collaborative filtering}

\author{Soudabeh Barghi, Lalet Scaria, Ali Salari, Tristan Glatard\\
  Department of Computer Science and Software Engineering\\ Concordia University, Montreal, Quebec, Canada}

\maketitle

\begin{abstract}
Evaluating the 
computational reproducibility of data analysis pipelines has become a 
critical issue. It is, however, a cumbersome 
process for analyses that involve data from large populations of 
subjects, due to their computational and storage requirements. 
We present a method to predict the computational 
reproducibility of data analysis pipelines in large population studies. 
We formulate the problem as a collaborative filtering process, with 
constraints on the construction of the training set. We propose 6 
different strategies to build the training set, which we evaluate on 2 
datasets, a synthetic one modeling a population with a growing number 
of subject types, and a real one obtained with neuroinformatics 
pipelines. Results show that one sampling method, ``Random File Numbers 
(Uniform)" is able to predict computational reproducibility with a good 
accuracy. We also analyze the relevance of including file and subject 
biases in the collaborative filtering model. We conclude that the 
proposed method is able to speed-up reproducibility 
evaluations substantially, with a reduced accuracy loss.
\end{abstract}

\section{Introduction}

Computational reproducibility, the ability to recompute analyses 
over time and 
space~\cite{peng2011reproducible}, has become a critical component of 
scientific methodology as many researchers acknowledge the existence of 
a reproducibility crisis~\cite{baker2016there}. Among other factors, infrastructural 
characteristics play an important role in making experiments 
reproducible. For 
instance, in 
neuroinformatics, our primary field of interest, studies have 
shown the effect of the 
operating system on computational 
results~\cite{gronenschild2012effects, glatard2015reproducibility}.
However, conducting such reproducibility studies at scale is cumbersome 
due to the computational and storage requirements of analysis 
pipelines.

Neuroinformatics pipelines are generally iterated on data coming from 10 to 
1,000 subjects, possibly with subtle input parameter variations to 
adjust specific data acquisition conditions. Subject data often capture 
anatomical and functional characteristics of their brain, for instance through
Magnetic Resonance Imaging (MRI) or Electroencephalography (EEG). The number of input files 
associated with a subject may vary, and these files may also be of 
different sizes. Typical processing times range from 15 minutes to 
15 hours per subject, with input ranging from 100~MB to 15~GB per subject, and 
outputs ranging from 1~GB to 500~GB.

The reproducibility of a given pipeline may vary across subjects, for 
instance due to different pipeline branches being executed depending on 
the input data content. As an example, in 
the public database released by the Human Connectome 
Project~\cite{van2013wu}, subjects may have one or two anatomical 
images of each modality; when two images are present, they are
aligned together and averaged. Acquisition artifacts, for instance due to motion,
may also trigger corrections not otherwise required, and some 
branches of the pipeline may be executed only for specific 
acquisition 
parameters. These remarks are 
consistent with recent findings showing that variations in the 
results of a functional MRI analysis are dependent on the dataset being 
analyzed~\cite{bowring2018exploring}. To capture such variations, 
reproducibility evaluations need to be conducted on many subjects, 
which is unwieldy. 

Our goal is to predict the outcome of reproducibility evaluations in a 
large population of subjects, from a reduced set of pipeline 
executions. More precisely, we aim at predicting whether a particular 
file produced by an analysis pipeline will be identical across 
execution conditions, or if it will contain reproducibility 
\emph{errors}. We approach the problem from the point of view of 
collaborative filtering, inspired by works that applied 
this method outside of its initial application domain, for instance 
~\cite{feng2013efficient}. The main issue, and originality, of our 
problem lies in the fact that the training set cannot be arbitrarily 
sampled from the utility matrix defining the collaborative filtering problem. Instead, the training set has to 
respect time constraints imposed by the file creation order during 
pipeline executions.

This paper makes the following contributions:
\begin{enumerate}
\item We model reproducibility evaluations in data processing pipelines as a collaborative filtering problem.
\item We propose strategies to sample the training set under time constraints.
\item We evaluate and compare our sampling strategies on synthetic and real datasets.
\end{enumerate}
The problem formulation, collaborative filtering technique, and 
proposed sampling strategies for the training set are presented in 
Section~\ref{sec:methods}. The datasets are described in 
Section~\ref{sec:datasets}, and experimental results are in 
Section~\ref{sec:results}.
Finally, we conclude on the best sampling method to use, and on the 
impact, limitations and generalizability of the results.

\section{Method}
\label{sec:methods}
\subsection{Problem formulation}

The pipeline to be evaluated is represented by a matrix $U$ of size 
$N_f \times N_s$, where the $N_s$ columns represent data coming from 
different subjects, and the $N_f$ rows represent the files produced by 
the pipeline. While subjects are not ordered, files are, for 
instance from their last modification time in a sequential execution, 
and we assume that this order is consistent across subjects. $U_{i, j}$ 
measures the reproducibility of file $i$ produced during the processing 
of subject $j$ in two conditions, for instance two different 
operating systems. In our experiments, we restrict $U_{i,j}$ to be 
boolean, but our methods can be applied for real values as well. 

Our goal is to predict the test set $\mathbb{T'}$ of missing values of 
$U$ from a training set $\mathbb{T}$ of known ones, where $\mathbb{T} 
\cap \mathbb{T'} = \emptyset$ and $\mathbb{T} \cup \mathbb{T'} = 
\{U_{i,j}\}$. In contrast with the traditional collaborative filtering 
problem, we have control over the construction of the training set. 
For instance, we can choose to include only files produced by specific 
subjects, or the first files produced by the processing of every 
subject. Moreover, the construction of the training set is constrained by 
the 
order of the matrix rows, which is formalized as follows: 
\begin{equation}
\begin{array}{l}
\forall (i, j, k) \in \llbracket 1, N_f \rrbracket \times \llbracket 1, N_f \rrbracket \times \llbracket 1, N_s \rrbracket, \\
 \left( U_{i,k} \in \mathbb{T} \quad \mathrm{and} \quad U_{j,k} \in \mathbb{T'} \right) \Rightarrow i < j. \label{eq:time}
 \end{array}
\end{equation}

Our problem is the following:
\begin{quote}
Given a training ratio $\alpha$, find a subset $\mathbb{T}$ of 
$\{U_{i,j}\}$ of size $\alpha N_f N_s$ such that (1) $\mathbb{T}$ and 
$\mathbb{T'}$ respect Equation~\ref{eq:time}, and (2) $\mathbb{T'}$ can 
be predicted from $\mathbb{T}$ with high accuracy.
\end{quote}
The sampling of the training set will be described in Section~\ref{sec:training}.
The collaborative filtering techniques used for the predictions are reported hereafter.

\subsection{Collaborative filtering}

Collaborative filtering is a technique to predict unknown values of a 
matrix called ``utility matrix" from the known ones. Traditionally, the 
matrix represents the ratings of items, represented in columns, by 
users, represented in rows. Ratings might be explicit, when users 
provide ratings through a dedicated system, or implicit, when users' 
behaviors are analyzed to estimate their preferences. An 
overview of collaborative filtering is provided 
in~\cite{leskovec2014mining}. In our context, the utility matrix is the
matrix $U$ described previously.

Several methods have been proposed to implement collaborative 
filtering. Item-item collaborative filtering~\cite{breese1998empirical, linden2003amazon} predicts 
the rating of item $i$ by user $u$ from the ratings of items similar to 
$i$ by user $u$. Likewise, user-user collaborative 
filtering~\cite{breese1998empirical} predicts the rating of item $i$ by user $u$ 
from the ratings of item $i$ by users similar to $u$. Both methods
have been used extensively for e-commerce applications.

A third class of methods, which is the one that we will use, is based 
on the factorization of the utility matrix to estimate latent factors 
along which users and items are represented. This method is described 
in~\cite{koren2009matrix} and became famous as it contributed to 
winning the 
Netflix prize in 2009. To summarize, the method aims at finding $q_i$ 
and $p_u$ vectors of $\mathbb{R}^f$, where $f$ is the number of latent factors, such that:
\begin{equation*}
r_{ui} = q_i^Tp_u,
\end{equation*}
where $r_{ui}$ is the rating of item $i$ by user $u$. In practice, the optimization
involves a regularization term to avoid overfiting particular users or items. The method
finds $q_i$ and $p_u$ that minimize the following objective, where $\lambda$ is the regularization parameter and $\mathbb{T}$ is the training set:
\begin{equation*}
\sum_{(i,u) \in \mathbb{T}}\left( r_{ui} - q_i^Tp_u\right)^2+\lambda \left( \|{q_i}\|^2 + \|{p_u}\|^2\right).
\end{equation*}

It is also common to include user biases $b_u$ and item biases $b_i$ in 
the optimization, defined as the average deviation of user $u$ and item 
$i$ to the global average $\mu$. The problem is then to find $q_i$ and $p_u$ that minimize the 
following objective:
\begin{equation*}
\sum_{(i,u) \in \mathbb{T}}\left( r_{ui} - \mu - b_u - b_i - q_i^Tp_u\right)^2+\lambda \left( \|{q_i}\|^2 + \|{p_u}\|^2 + b_u^2 + b_i^2\right)
\end{equation*}

Stochastic gradient descent and alternating least squares (ALS) are 
often used as optimization techniques. In the remainder, we use ALS as 
implemented in Apache Spark's MLLib version 2.3.0. We also round predictions 
to the nearest integer to obtain binary values.

\subsection{Sampling of the training set}

\label{sec:training}

As explained before, the training set in our problem cannot be 
constructed by unconstrained random sampling of the matrix. Instead, 
the training and test sets have to comply to Equation~\ref{eq:time}. To 
address this issue, we investigated the following sampling techniques, illustrated in
Figure~\ref{fig:sampling}. 

\subsubsection{Complete Columns 
(Figure~\ref{fig:Columns-Sample-Training-set})} The training set is 
sampled by randomly selecting complete columns in the utility matrix, 
that is, complete subject executions. The last selected column might be 
incomplete to meet the exact training ratio. This method respects the 
time constraints. It corresponds to a situation where the collaborative 
filtering method will predict the reproducibility of the subjects in 
the test set from the subjects in the training set. 

\begin{figure}[h!]
        \centering
        \begin{subfigure}[b]{0.45\columnwidth}
                  \includegraphics[width=\columnwidth]{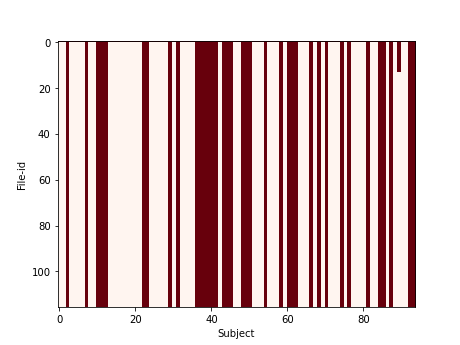}
                  \caption{Complete Columns}
                  \label{fig:Columns-Sample-Training-set}
        \end{subfigure}
        \begin{subfigure}[b]{0.45\columnwidth}
                  \includegraphics[width=\columnwidth]{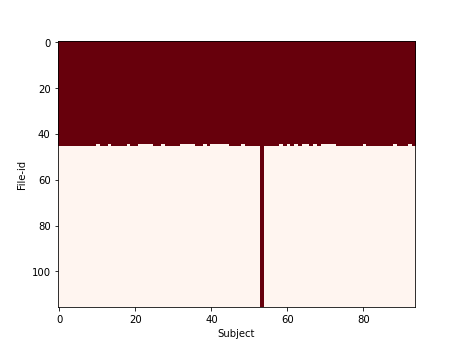}
                  \caption{Complete Rows}
                  \label{fig:Rows-Sample-Training-set}
        \end{subfigure}
        \begin{subfigure}[b]{0.45\columnwidth}
                 \includegraphics[width=\columnwidth]{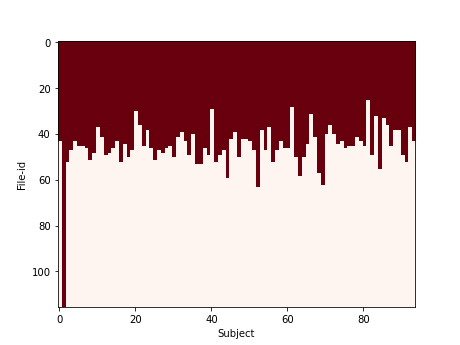}
                \caption{Random Subjects (RS)}
                  \label{fig:Uniform-S-Sample-Training-set}
        \end{subfigure}
        \begin{subfigure}[b]{0.45\columnwidth}
                  \includegraphics[width=\columnwidth]{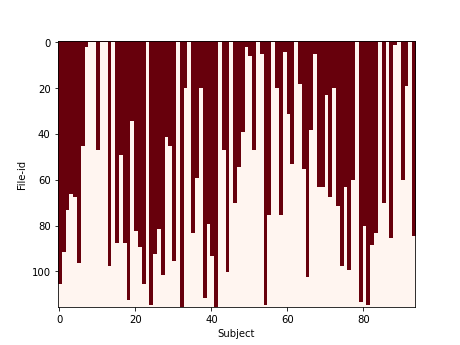}
                  \caption{RFNU}
                  \label{fig:Diagonal-Sample-Training-set}
        \end{subfigure}
        \begin{subfigure}[b]{0.45\columnwidth}
                  \includegraphics[width=\columnwidth]{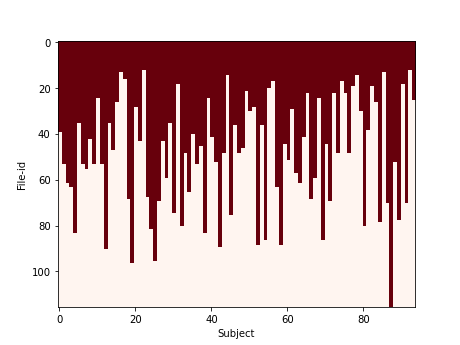}
                  \caption{RFNTL}
                  \label{fig:triangular-L-Sample-Training-set}
        \end{subfigure}
        \begin{subfigure}[b]{0.45\columnwidth}
                  \includegraphics[width=\columnwidth]{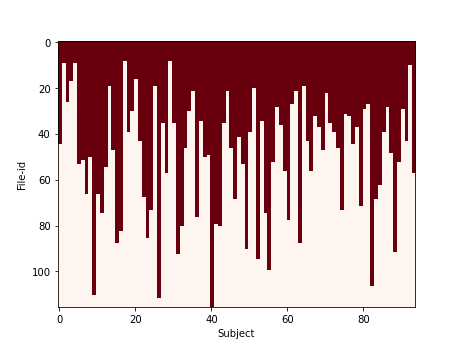}
                  \caption{RFNTS}
                  \label{fig:triangular-S-Sample-Training-set}
        \end{subfigure}
        \begin{subfigure}[b]{\columnwidth}
        \centering
        \includegraphics[width=0.45\columnwidth]{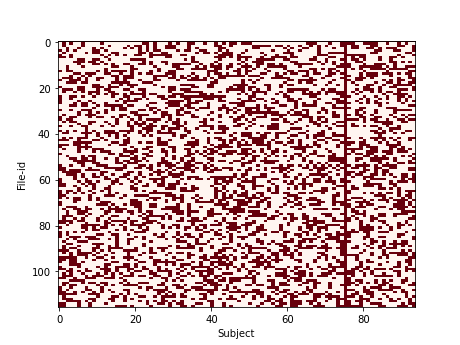}
                  \caption{Random Unreal}
                  \label{fig:Random-Unreal-Sample-Training-set}
        \end{subfigure}
        \caption{Training sets for different sampling methods, $\alpha=0.4$ (dark elements are in the training set).}
        \label{fig:sampling}
\end{figure}

\subsubsection{Complete Rows (Figure~\ref{fig:Rows-Sample-Training-set})} 
The training set is sampled by selecting complete rows in the utility 
matrix, that is, the first files produced by every execution. The last 
selected row might be incomplete to meet the exact training ratio. This 
method respects the time constraints. It corresponds to a situation 
where the processing of all the subjects is launched and interrupted 
before the execution is complete. The collaborative filtering method 
will then predict the reproducibility of the remaining files.

\subsubsection{Random Subjects -- RS 
(Figure~\ref{fig:Uniform-S-Sample-Training-set})} This method builds 
the training set by selecting the files from random subjects until the 
training ratio is reached. The file selected in a subject is the file 
with the lowest index in this subject that has not been already 
selected in the training set, which respects the time constraints.

\subsubsection{Random File Numbers (Uniform) -- RFNU (Figure~\ref{fig:Diagonal-Sample-Training-set})}
The number of files selected for every subject is randomly selected in
a uniform distribution $U(\textit{a},\textit{b})$, where \textit{b} is set to the total
number of files $N_{f}$ and \textit{a} is set according to training ratio $\alpha$ as follows:
\[
  \begin{cases}
          \textit{a} = 0      & \text{if $\alpha \leq 0.5$ }\\
          
          \textit{a} = (2\alpha - 1) N_{f} & \text{if $\alpha > 0.5$}
  \end{cases}
\]
 For $\alpha \leq 0.5$, we ensure that the average number of selected 
 files by subject is $\alpha N_f$ by sampling the number of 
 files in every subject from $U(0,N_f)$ with 
 probability $2\alpha$, and setting it to 0 otherwise. For $\alpha > 
 0.5$, this is ensured by the value of $a$, which leads the 
 average of U(a, b) to be $\alpha N_f$.

\subsubsection{Random File Numbers (Triangular) -- RFNT}
\begin{figure}
\centering
\includegraphics[width=0.5\columnwidth]{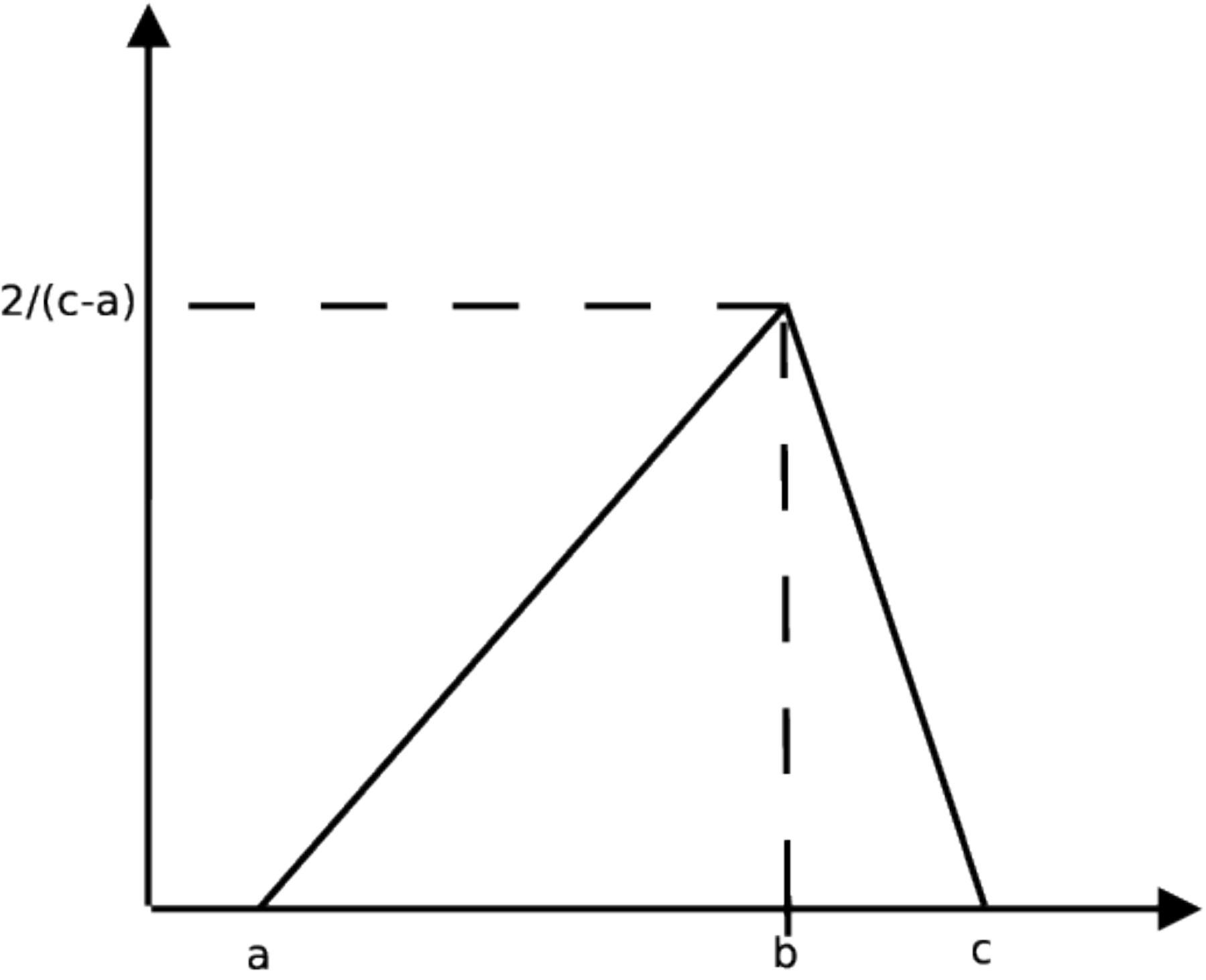}
\caption{Triangular distribution $T(a, b, c)$}
\label{fig:triangular}
\end{figure}

The number of files selected for every subject is randomly selected in
a triangular distribution $T(a, b, c)$ as in Figure~\ref{fig:triangular}. The mean of the distribution is 
$\frac{a+b+c}{3}$. We set \textit{c} to $N_{f}$ and we set 
\textit{a} and \textit{b} with the following two approaches.

\paragraph{Largest a (RFNTL, 
Figure~\ref{fig:triangular-L-Sample-Training-set})} a is set to the 
largest possible value, i.e., b, and b is set accordingly to ensure 
that the average of the distribution is $\alpha N_{f}$. Two cases occur:
\begin{itemize}
\item When $\alpha > 1/3$:
\[ \textit{a}=\textit{b}=\frac{3\alpha-1}{2}N_{f}
\]
\item When $\alpha \leq 1/3$: $a=b=0$, the number of files selected 
in every subject is sampled from $T(0, 0, N_f)$ with probability 
$3\alpha$ and set to 0 otherwise.
\end{itemize}

\paragraph{Smallest a (RFNTS, 
Figure~\ref{fig:triangular-S-Sample-Training-set})} a is set to the 
smallest possible value, i.e., 0, and b is set accordingly to ensure that the average of the
distribution is $\alpha N_{f}$. Three cases occur:
\begin{itemize}
\item When $\alpha < 1/3$: $a=b=0$, the number of files 
selected in every subject is sampled from $T(0, 0, N_f)$ with 
probability $3\alpha$ and set to 0 otherwise.
\item When $1/3 \leq \alpha \leq 2/3$:
\[
          \textit{a} = 0   \quad ; \quad
          \textit{b}=N_{f}(3\alpha-1)
\]
\item When $\alpha \geq 2/3$: $b=N_f$, the number of files selected in 
every subject is sampled from $T(0, N_f, N_f)$ with probability 
$3(1-\alpha)$ 
and set to $N_f$ otherwise.
\end{itemize}

As illustrated in Figures~\ref{fig:triangular-L-Sample-Training-set} 
and~\ref{fig:triangular-S-Sample-Training-set}, the point of the 
RFNTL method is to guarantee that, for large enough values of 
$\alpha$, all subjects will have at least a few files in the training 
set (no empty column), which is not the case for RFNU or RFNTS.

\subsubsection{Random Unreal (Figure 
~\ref{fig:Random-Unreal-Sample-Training-set})} The training set is 
sampled in a random uniform way, regardless of the file creation 
times. This method does not respect the time constraints in 
Equation~\ref{eq:time}: it will be used as a baseline for comparison.

In each method, we also included the first row of the matrix (first 
 file of each subject) and a random column (all files of a 
random subject) to avoid cold start issues.

\section{Datasets}

\label{sec:datasets}

\subsection{Synthetic Dataset}

\begin{figure}
\centering
        \begin{subfigure}[b]{0.45\columnwidth}
                  \includegraphics[width=\columnwidth]{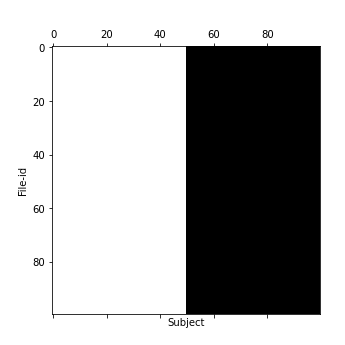}
                  \caption{2 types}
        \end{subfigure}
        \begin{subfigure}[b]{0.45\columnwidth}
                  \includegraphics[width=\columnwidth]{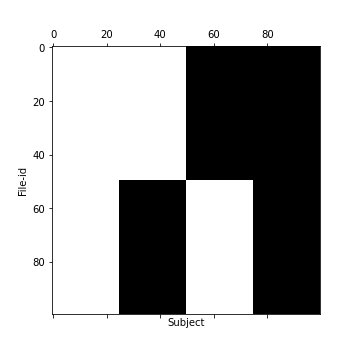}
                  \caption{4 types}
        \end{subfigure}
                \begin{subfigure}[b]{0.45\columnwidth}
                  \includegraphics[width=\columnwidth]{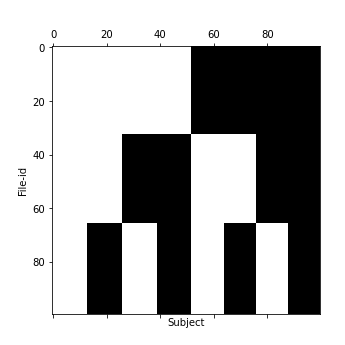}
                  \caption{8 types}
        \end{subfigure}
                \begin{subfigure}[b]{0.45\columnwidth}
                  \includegraphics[width=\columnwidth]{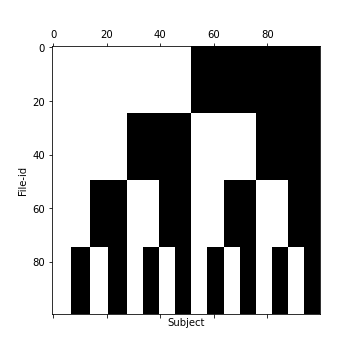}
                  \caption{16 types}
        \end{subfigure}
                \begin{subfigure}[b]{0.45\columnwidth}
                  \includegraphics[width=\columnwidth]{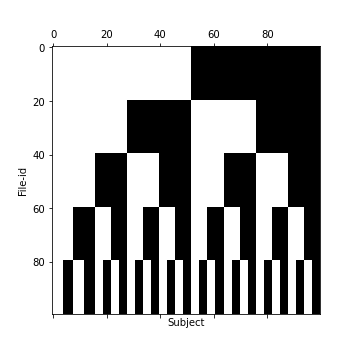}
                  \caption{32 types}
        \end{subfigure}
                \begin{subfigure}[b]{0.45\columnwidth}
                  \includegraphics[width=\columnwidth]{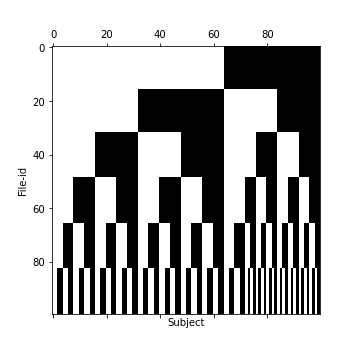}
                  \caption{64 types}
        \end{subfigure}
\caption{Synthetic reproducibility matrices. White cells denote reproducibility errors.}
\label{fig:synthetic-data}
\end{figure}

We generated synthetic matrices as shown in 
Figure~\ref{fig:synthetic-data}. Each matrix has 100 files and 100 
subjects of different \emph{types}. Subjects of the same type behave 
identically and all types contain the same number of subjects $\pm$ 1. 
Such a decomposition by subject type corresponds to a situation where 
different subjects may have data of different nature, as is the case in the
Human Connectome Project data~\cite{van2013wu} where not all the 
subjects have the same amount of images.

 For subjects of a given type, the matrix consists of 
 $log(n)$  blocks, where $n$ is the number of types. Blocks are 
 defined with all the possible variation patterns: some types do not 
 vary at all, while other ones vary between every block. Such 
 patterns are meant to mimic the logic of data 
processing pipelines: each block of files represents the files produced 
at a given stage of the pipeline, which may or may not contain 
reproducibility errors depending on the subject type.

\subsection{Real Dataset}

We
processed a set S of 94 subjects randomly selected in the S500 HCP 
release\footnote{\url{https://db.humanconnectome.org}} of the Human 
Connectome Project~\cite{glasser2013minimal}, in three execution 
conditions with different versions of the CentOS operating system 
(5.11, 6.8 and 7.2 -- referred as C5, C6 and C7), using the 
PreFreesurfer and Freesurfer pipelines described 
in~\cite{glasser2013minimal} and available on 
GitHub\footnote{\url{https://github.com/Washington-University/Pipelines/releases/tag/v3.19.0}}. 
For each pipeline, we identified the set F of files produced for all 
subjects in all conditions. For each condition pair and each pipeline, 
we computed a binary reproducibility matrix U of size $|F|\times|S|$, 
where $U_{i,j}$ is true if and only if file $i$ of subject $j$ was 
different in each condition. Rows of $U$ were ordered by ascending file 
modification time in a random subject in S.

\begin{figure}[h!]
  \centering
  \begin{subfigure}[b]{0.45\columnwidth}
        \includegraphics[width=\columnwidth]{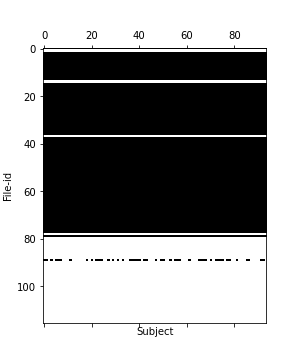}
  \caption{PFS, C5 vs C6}
  \end{subfigure}
  \begin{subfigure}[b]{0.45\columnwidth}
         \includegraphics[width=\columnwidth]{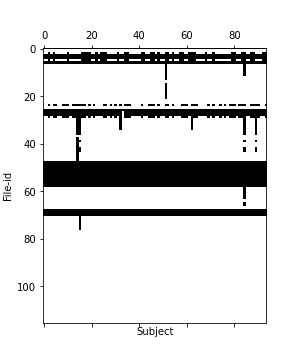}
  \caption{PFS, C5 vs C7}
  \end{subfigure}
  \begin{subfigure}[b]{0.45\columnwidth}
        \includegraphics[width=\columnwidth]{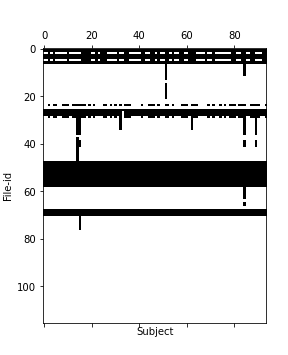}
  \caption{PFS, C6 vs C7}
  \end{subfigure}
  \begin{subfigure}[b]{0.45\columnwidth}
        \includegraphics[width=\columnwidth]{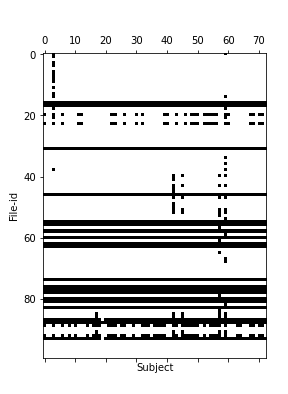}
  \caption{FS (100 files), C6 vs C7}
  \end{subfigure}
  \caption{Utility matrices for PreFreesurfer (PFS) and Freesurfer (FS). White cells denote reproducibility errors.}
    \label{fig:utility-matrices}
\end{figure}

Figure~\ref{fig:utility-matrices} shows the matrices
for PreFreesurfer and Freesufer. The
reproducibility of these pipelines varies across subjects,
but most files behave
consistently across all subjects, leading to complete black or white
lines.

\section{Results}
\label{sec:results}
We conducted two experiments for each reproducibility matrix, to evaluate the performance of our 
predictions using (1) ALS without biases, and (2) ALS with subject and file biases. We 
compare the performance of our sampling methods to (1) a dummy 
classifier that always predicts the value in the majority class and (2) 
the Random Unreal method, used as the baseline sampling technique. All 
reported values are averages over 5 repetitions. Due to sampling 
issues, it is possible that the actual training ratio obtained with 
some of the sampling methods does not exactly
match the target one. We checked that the difference between the target 
and real training ratios was lower than 0.01. We used Spark's ALS model 
as available in \texttt{pyspark.ml}, with 50 factors, a 
regularization parameter of 0.01, a maximum of 5 iterations and non 
negative constraints set to true. The code and data used to obtain the results are available through GitHub at
\url{https://github.com/big-data-lab-team/paper-reproducibility-collaborative-filtering}.


\subsection{Accuracy on Synthetic Data}

\subsubsection{ALS without Bias}
\begin{figure*}
\begin{subfigure}[b]{\columnwidth}
        \includegraphics[width=0.8\columnwidth]{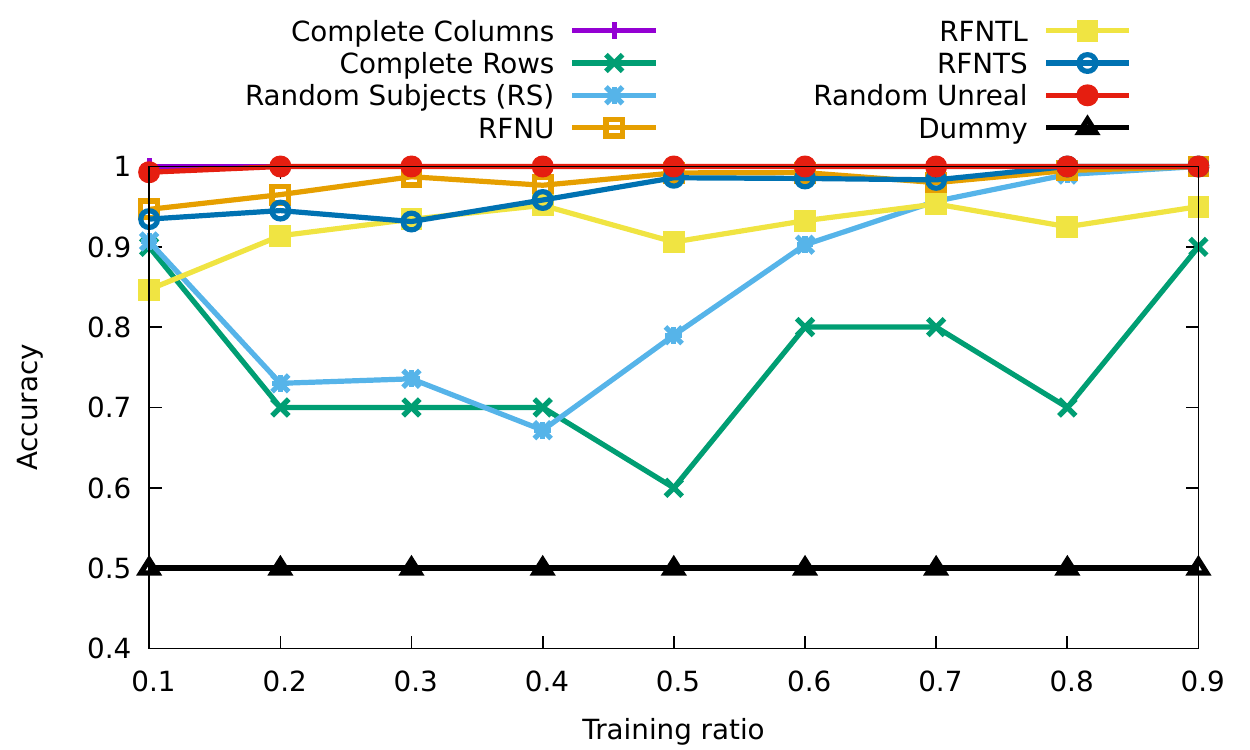}
        \caption{2 subject types}
\end{subfigure}
\begin{subfigure}[b]{\columnwidth}
        \includegraphics[width=0.8\columnwidth]{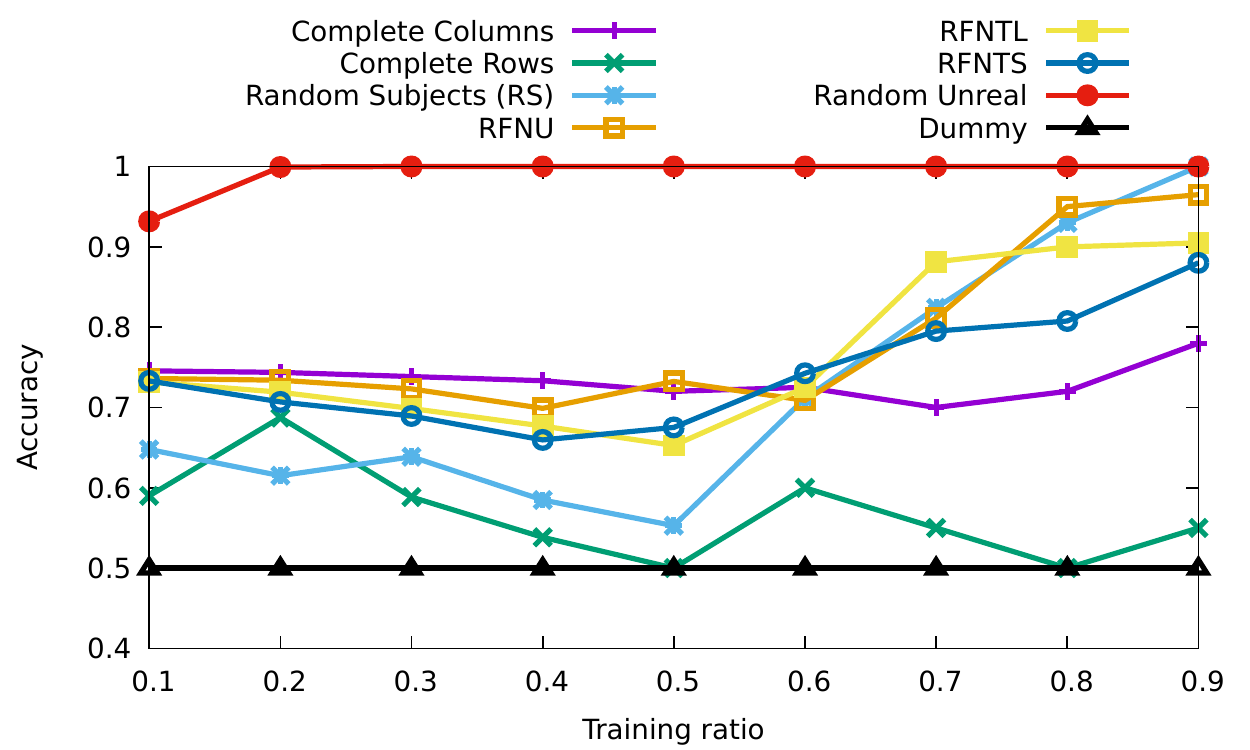}
        \caption{4 subject types}
\end{subfigure}
\begin{subfigure}[b]{\columnwidth}
        \includegraphics[width=0.8\columnwidth]{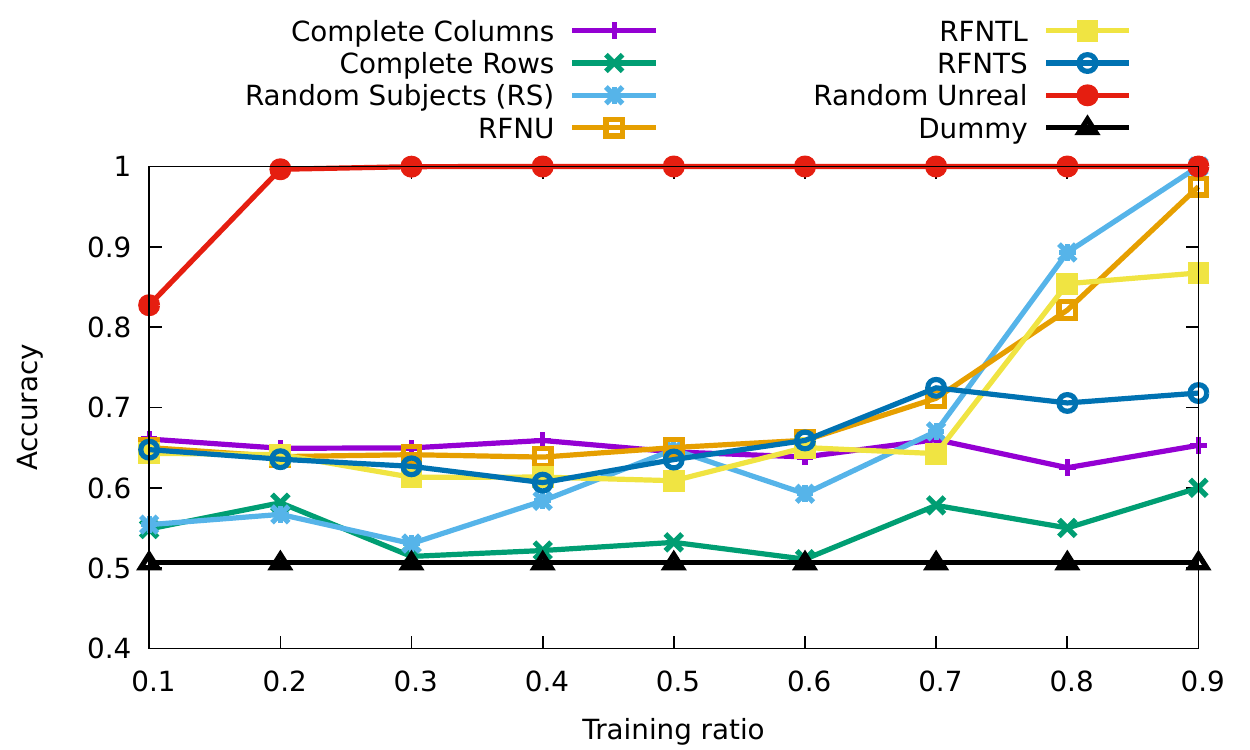}
        \caption{8 subject types}
\end{subfigure}
\begin{subfigure}[b]{\columnwidth}
        \includegraphics[width=0.8\columnwidth]{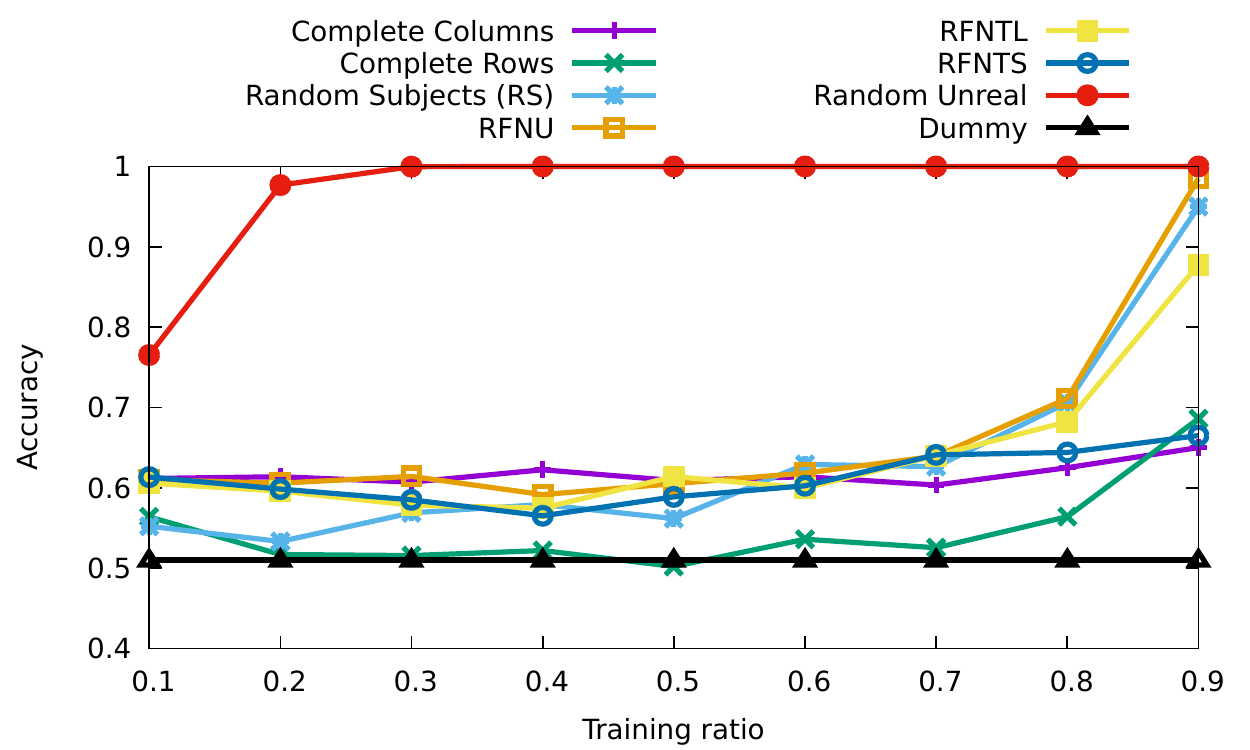}
        \caption{16 subject types}
\end{subfigure}
\begin{subfigure}[b]{\columnwidth}
        \includegraphics[width=0.8\columnwidth]{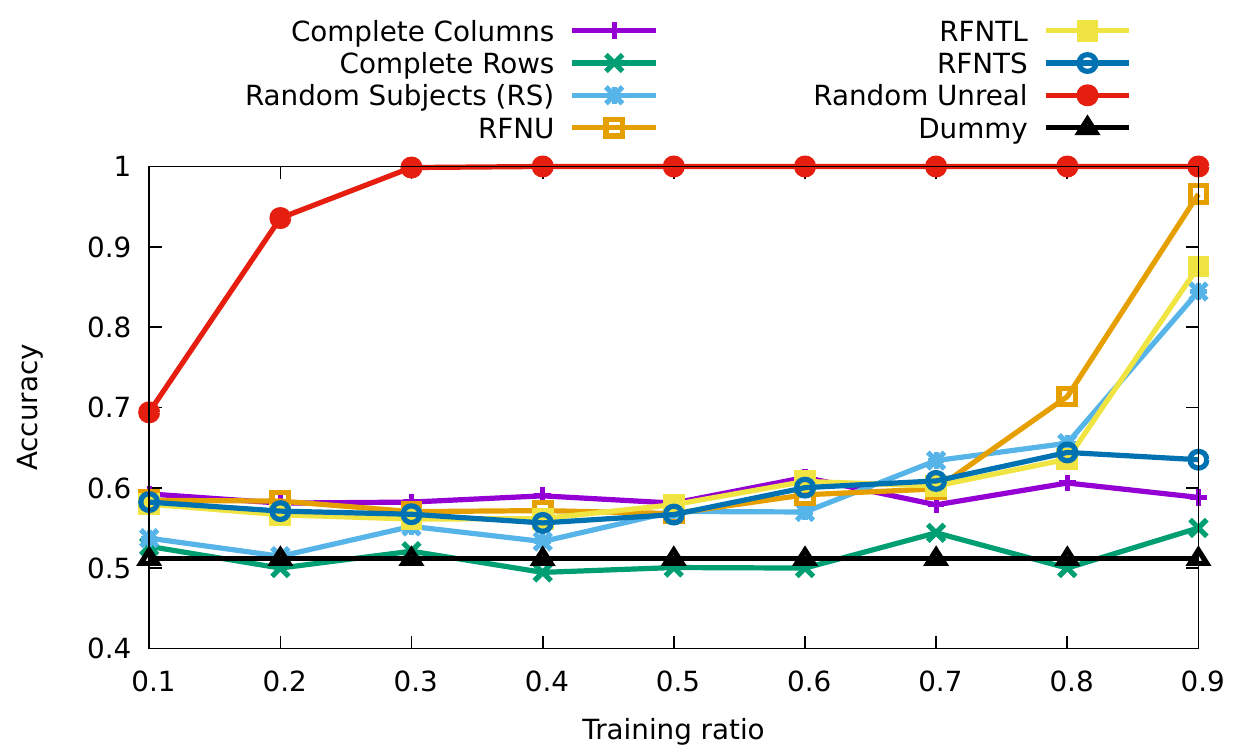}
        \caption{32 subject types}
\end{subfigure}
\hfill
\begin{subfigure}[b]{\columnwidth}
        \includegraphics[width=0.8\columnwidth]{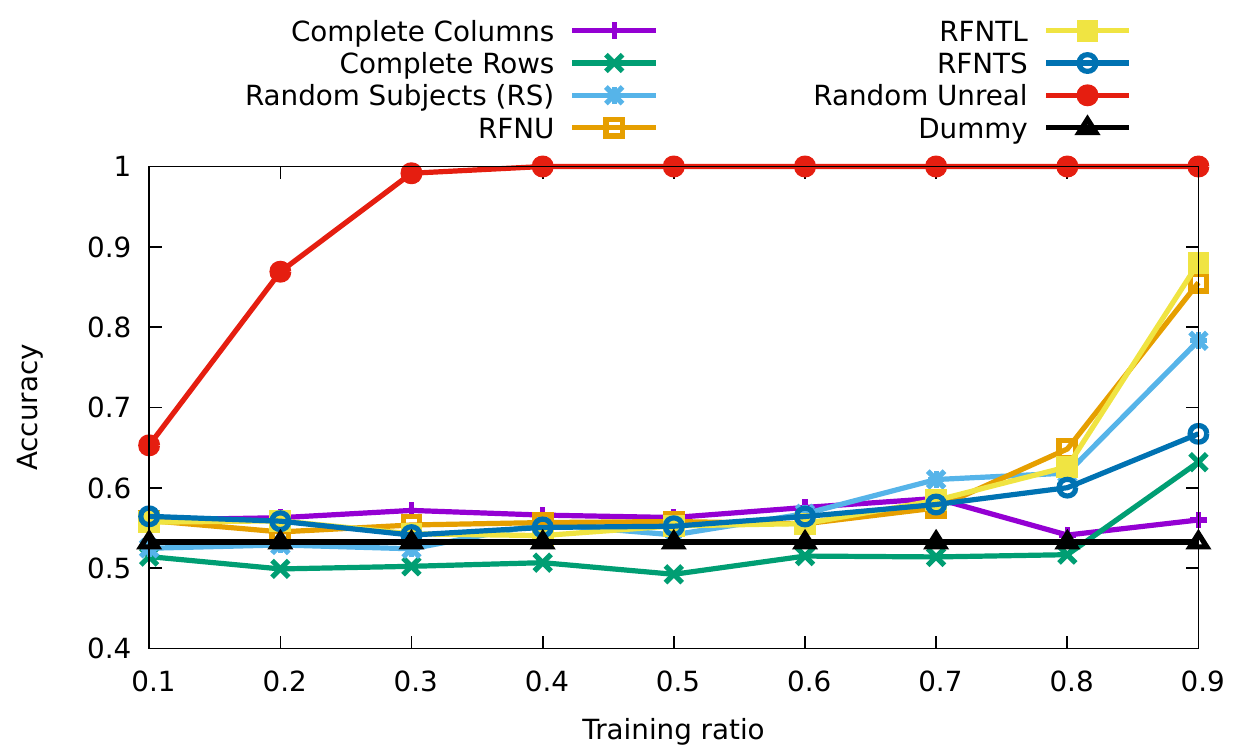}
        \caption{64 subject types}
\end{subfigure}
\caption{Accuracy results on synthetic data, ALS \emph{without} bias.}
\label{fig:results-synthetic}
\end{figure*}

 Figure~\ref{fig:results-synthetic} shows accuracy results for ALS 
 without bias. By construction, the accuracy of the dummy classifier is 
 close to 0.5 for all subject types. Random Unreal performs well 
 for all subject types, which confirms that ALS is working correctly. 
 All the other methods 
perform better than the dummy classifier, and their accuracy decreases 
as the number of subject types increases.
However, only 3 methods can provide accuracy values above 0.85 
for all subject types: Random Subjects (RS), RFNU and RFNTL. 
Surprisingly, RFNTS does not perform well for more than 2 subject 
types, even for $\alpha=0.9$.

\subsubsection{ALS with Bias}

\begin{figure*}
\begin{subfigure}[b]{\columnwidth}
        \includegraphics[width=0.8\columnwidth]{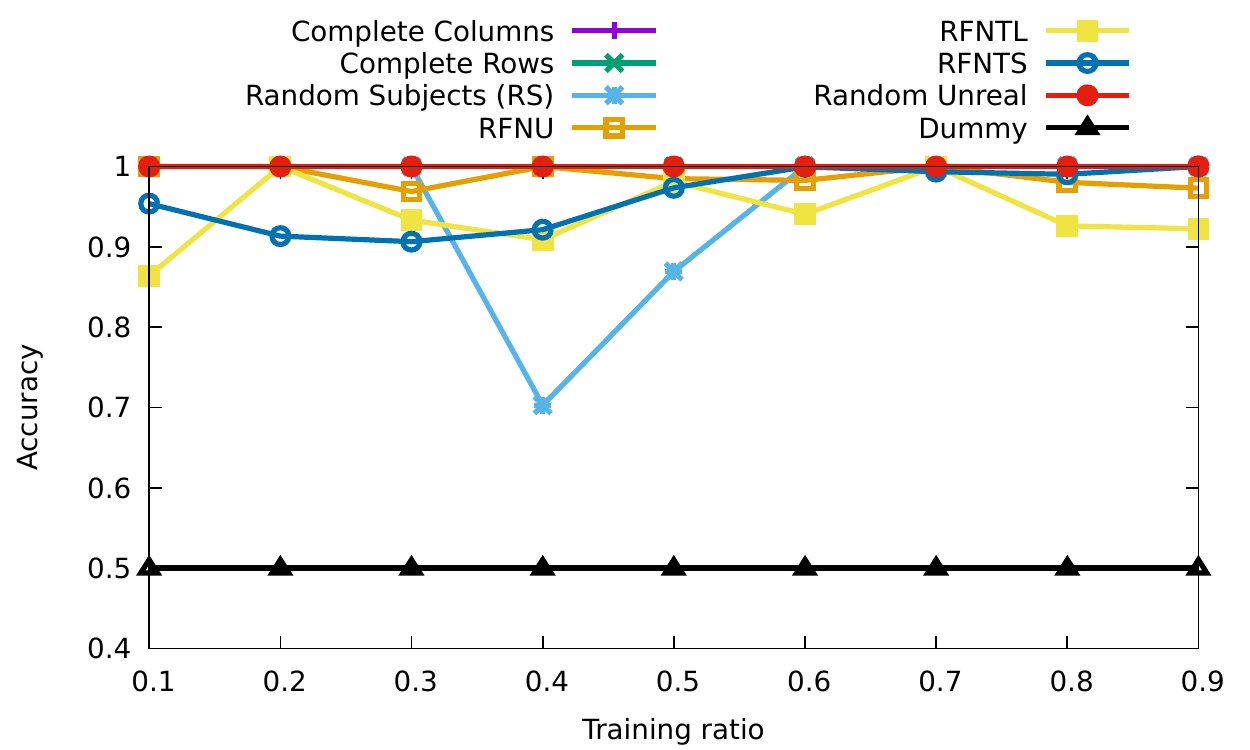}
        \caption{2 subject types}
\end{subfigure}
\begin{subfigure}[b]{\columnwidth}
        \includegraphics[width=0.8\columnwidth]{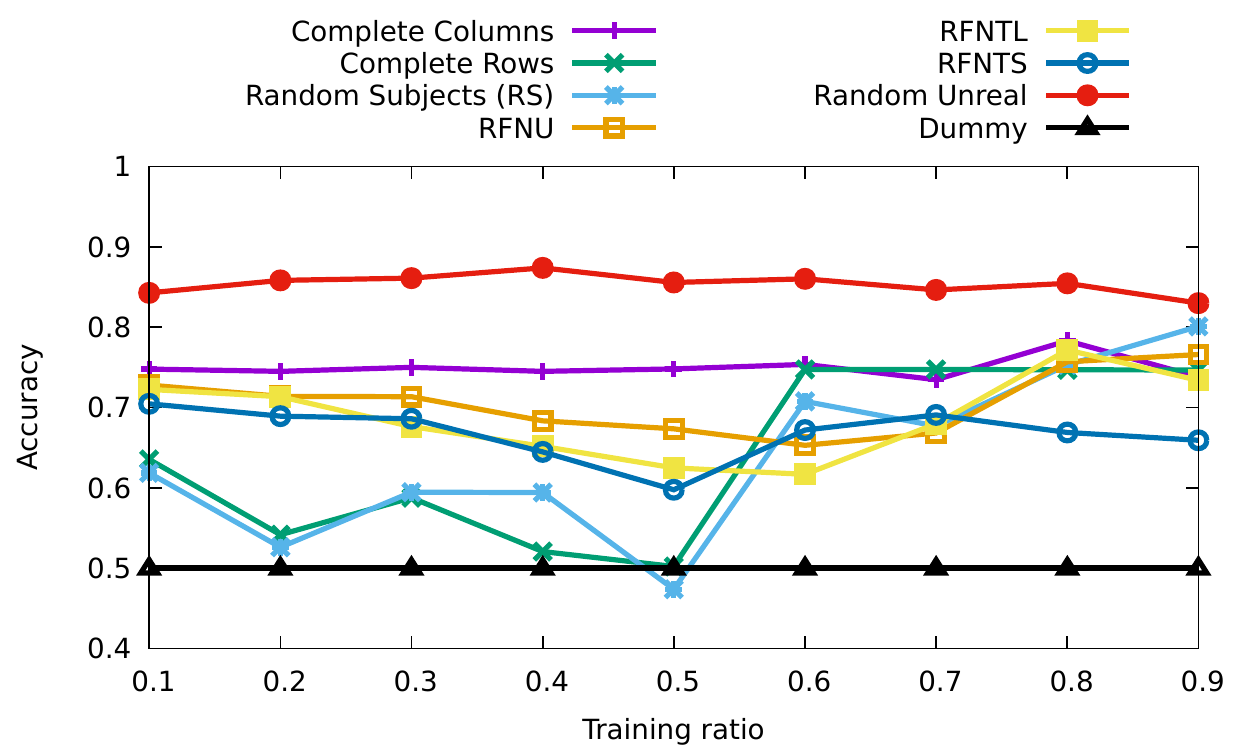}
        \caption{4 subject types}
\end{subfigure}
\begin{subfigure}[b]{\columnwidth}
        \includegraphics[width=0.8\columnwidth]{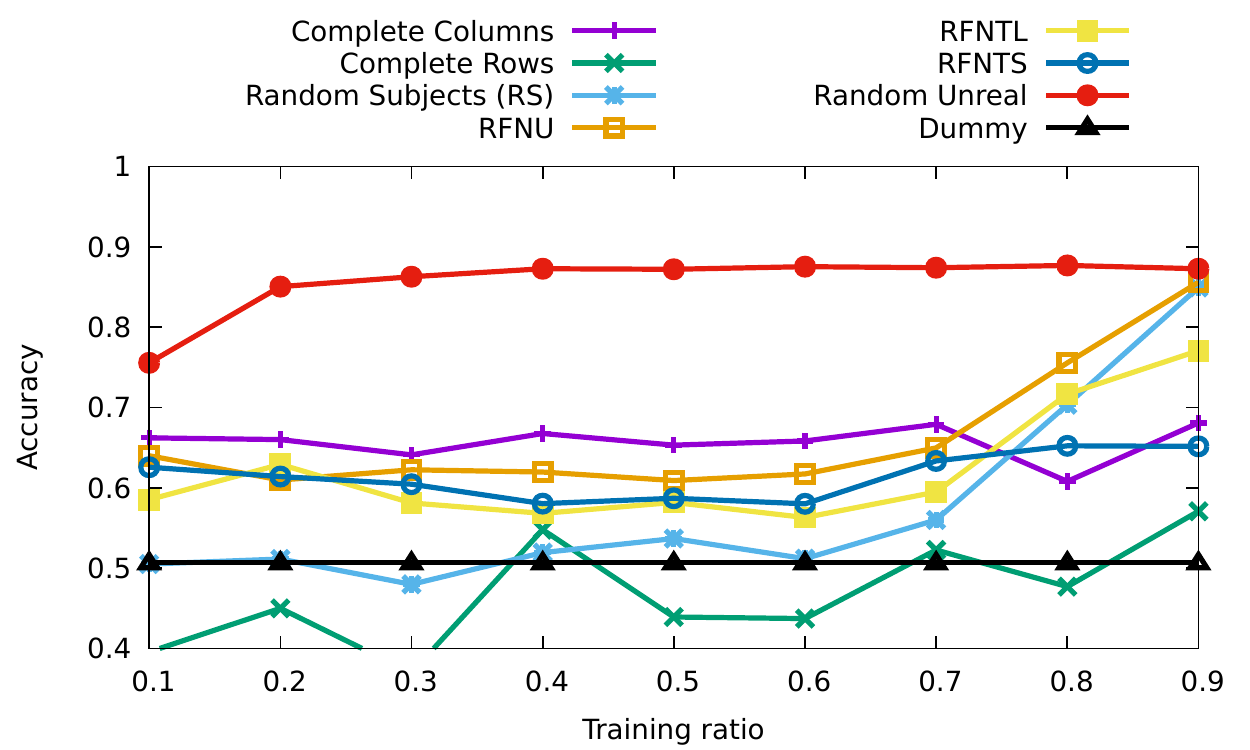}
        \caption{8 subject types}
\end{subfigure}
\begin{subfigure}[b]{\columnwidth}
        \includegraphics[width=0.8\columnwidth]{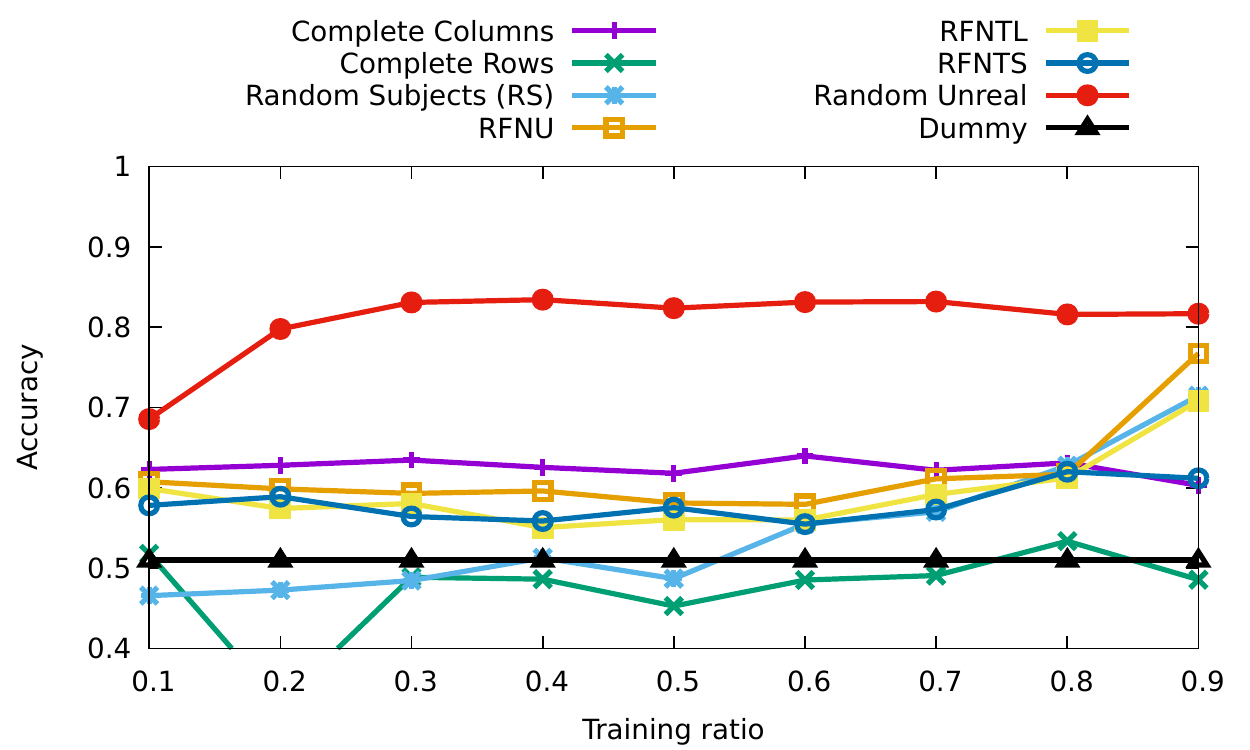}
        \caption{16 subject types}
\end{subfigure}
\begin{subfigure}[b]{\columnwidth}
        \includegraphics[width=0.8\columnwidth]{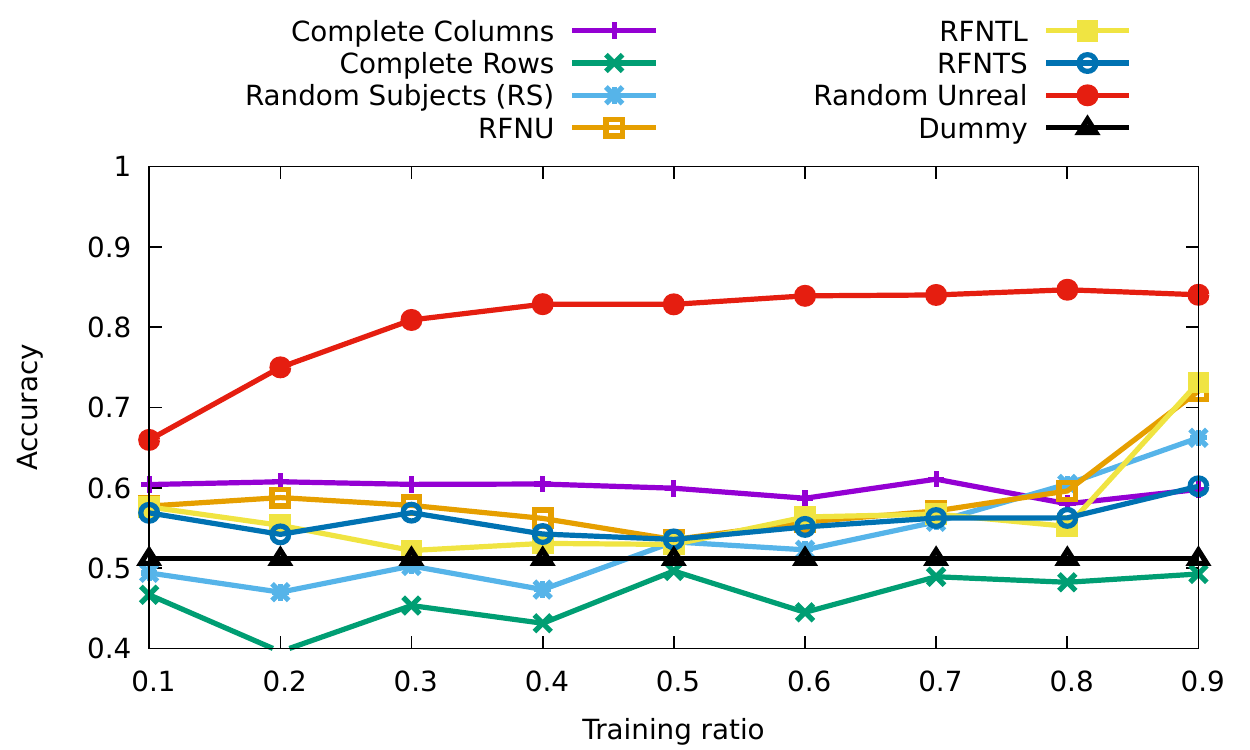}
        \caption{32 subject types}
\end{subfigure}
\hfill
\begin{subfigure}[b]{\columnwidth}
        \includegraphics[width=0.8\columnwidth]{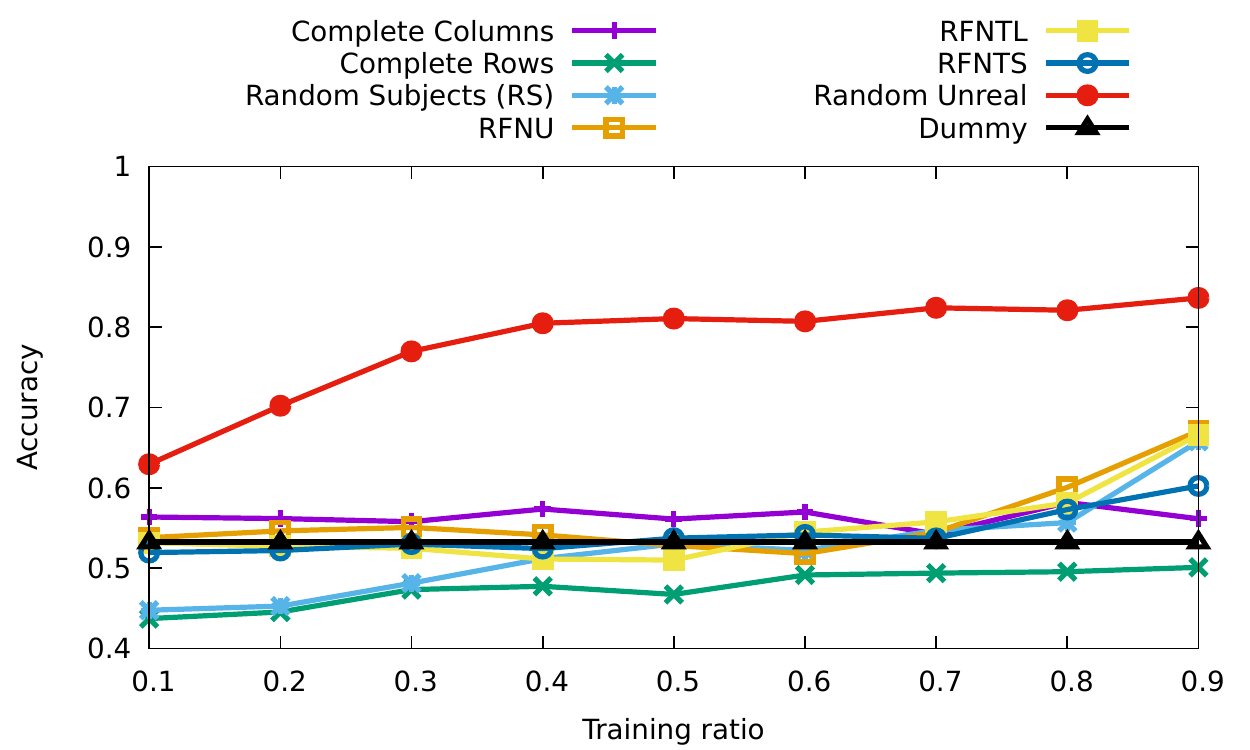}
        \caption{64 subject types}
\end{subfigure}
\caption{Accuracy results on synthetic data, ALS \emph{with} bias.}
\label{fig:results-synthetic-als-bias}
\end{figure*}

Figure~\ref{fig:results-synthetic-als-bias} shows accuracy results for 
ALS with bias. RS, RFNU and RFNTL remain the methods that best compare 
to Random Unreal, but their accuracy is much lower than without 
biases. This is presumably due to 
the fact that file biases are all close to 0.5. 
In such a situation, biases are detrimental and should not be included.

\subsection{Accuracy on  Real Data}

\subsubsection{ALS without bias}

\begin{figure*}
\begin{subfigure}[b]{\columnwidth}
        \includegraphics[width=0.8\columnwidth]{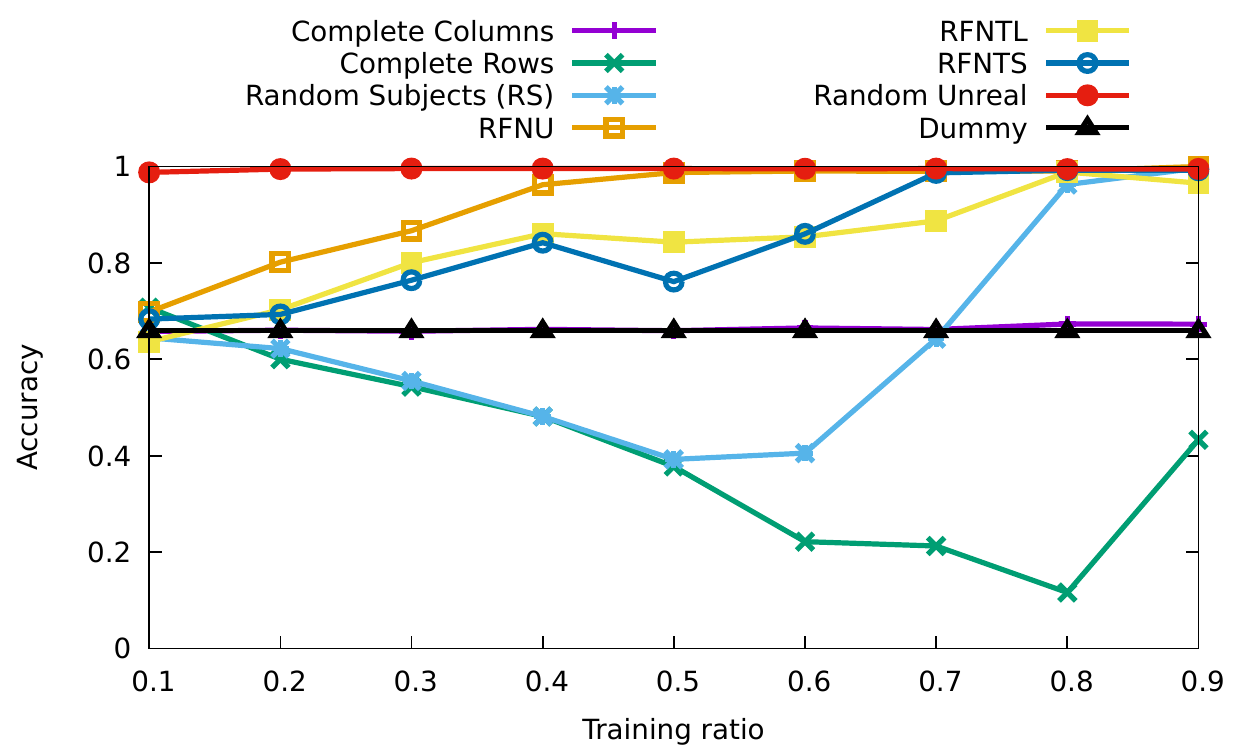}
        \caption{PFS, CentOS5 vs CentOS6}
\end{subfigure}
\begin{subfigure}[b]{\columnwidth}
        \includegraphics[width=0.8\columnwidth]{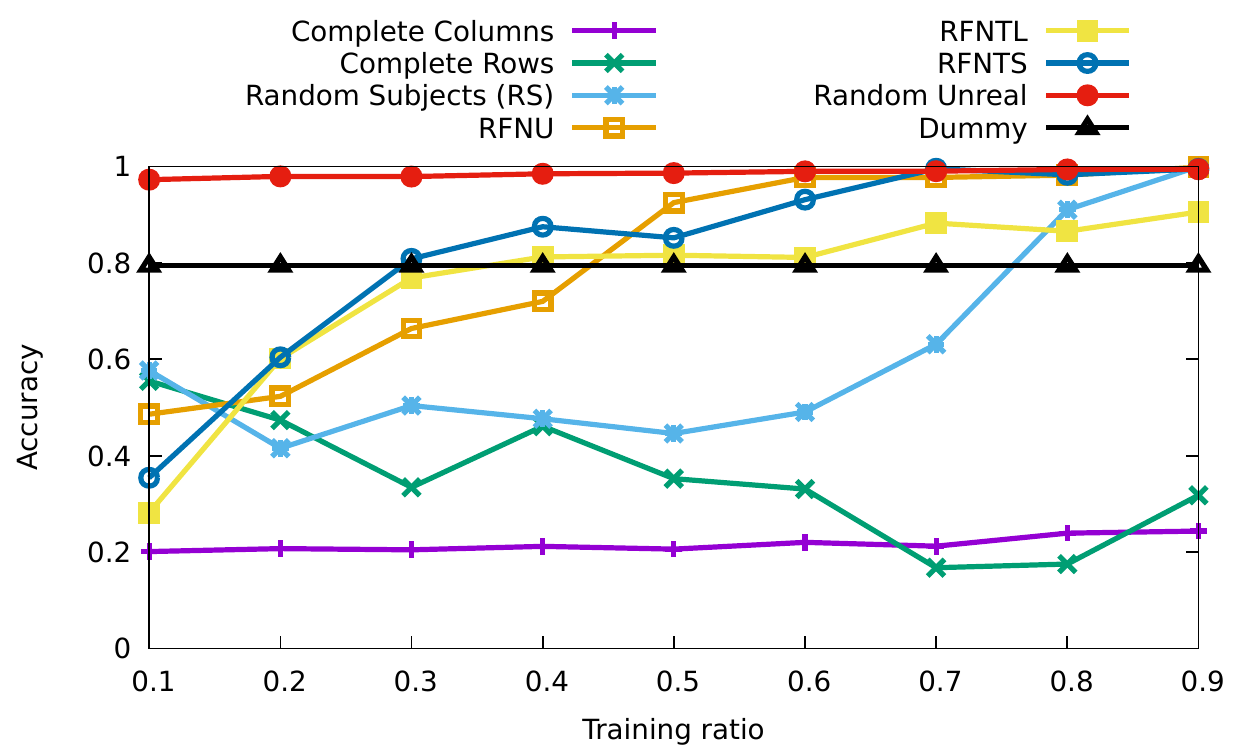}
        \caption{PFS, CentOS5 vs CentOS7}
\end{subfigure}
\begin{subfigure}[b]{\columnwidth}
        \includegraphics[width=0.8\columnwidth]{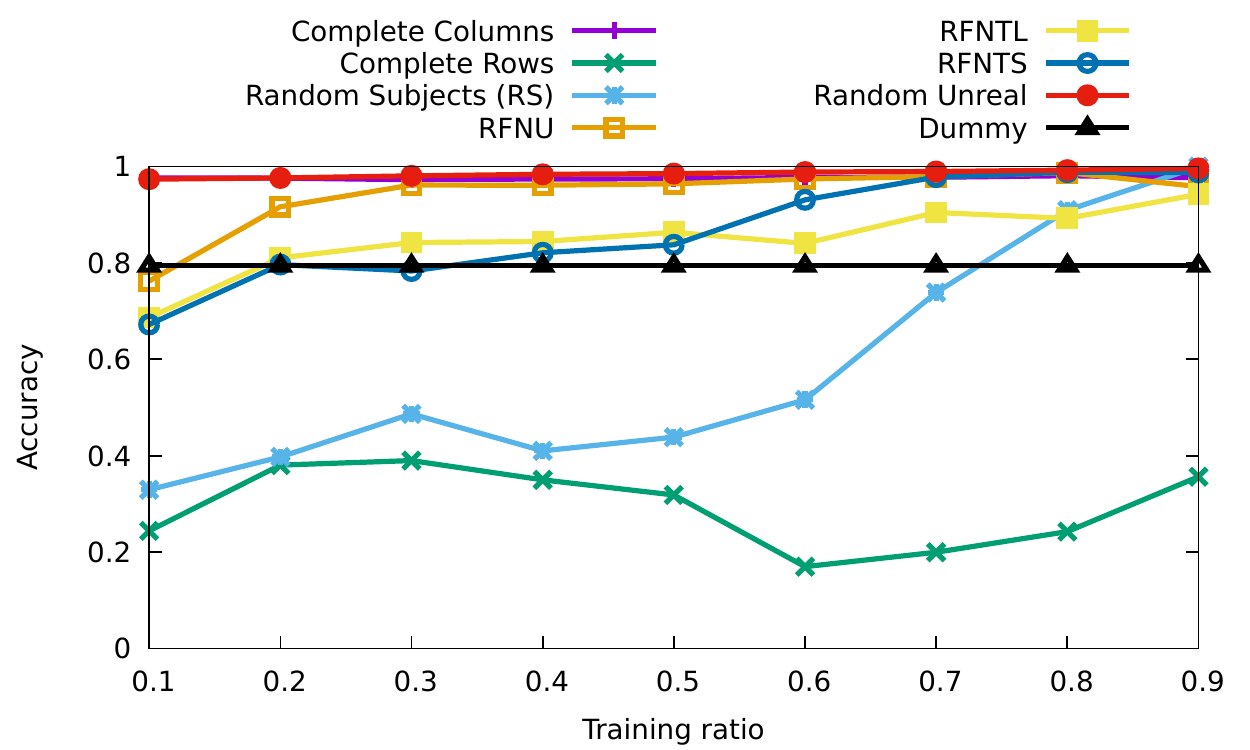}
        \caption{PFS, CentOS6 vs CentOS7}
\end{subfigure}\hfill
\begin{subfigure}[b]{\columnwidth}
        \includegraphics[width=0.8\columnwidth]{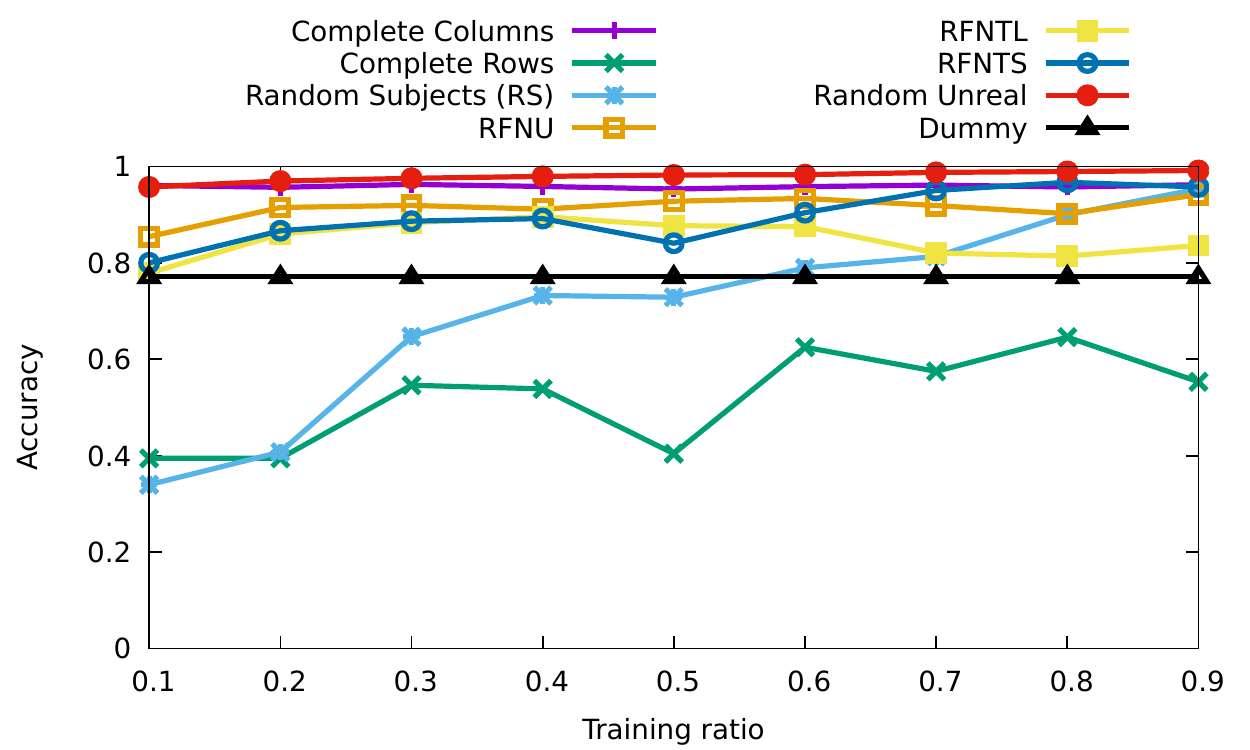}
        \caption{FS, CentOS6 vs CentOS7}
\end{subfigure}
\caption{Accuracy results on PreFresurfer (PFS) and Freesurfer (FS) data, ALS \emph{without} bias.}
\label{fig:results-real-als}
\end{figure*}

Figure~\ref{fig:results-real-als} shows the accuracy for ALS without bias. Among the methods that performed 
well in the synthetic dataset, RFNU performs the best, with an accuracy 
higher than 0.95 for all datasets when the training ratio is larger 
than 0.5. Complete rows and complete columns do not reach the accuracy of 
the dummy classifier. Random Unreal has an 
accuracy close to 1, which shows that collaborative filtering works 
well on this dataset too.

\subsubsection{ALS with bias}

\begin{figure*}
\begin{subfigure}[b]{\columnwidth}
        \includegraphics[width=0.8\columnwidth]{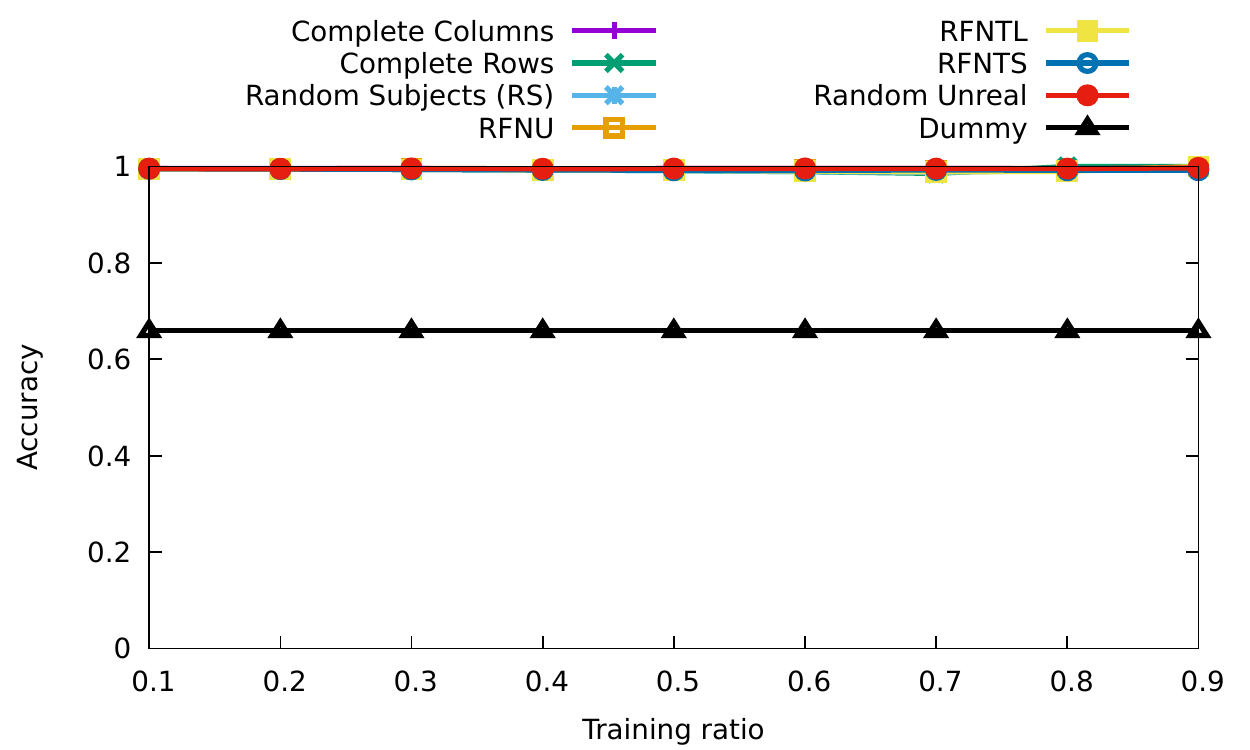}
        \caption{PFS, CentOS5 vs CentOS6}
\end{subfigure}
\begin{subfigure}[b]{\columnwidth}
        \includegraphics[width=0.8\columnwidth]{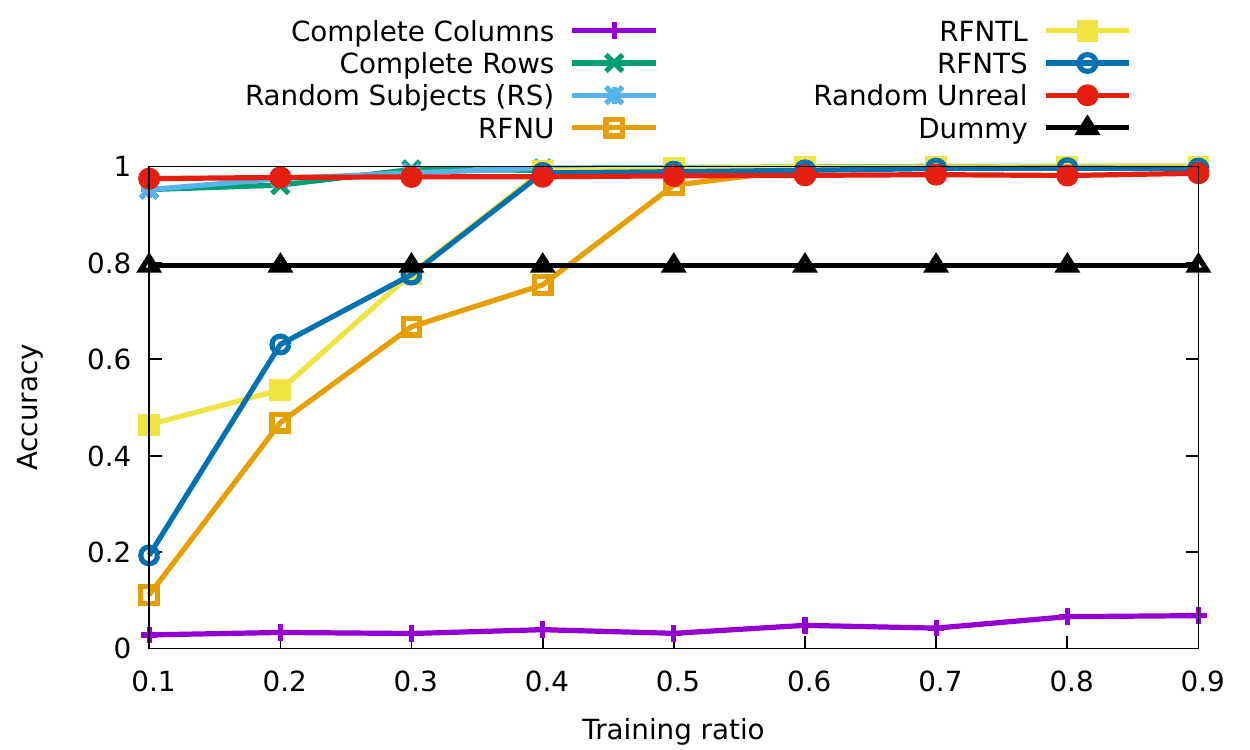}
        \caption{PFS, CentOS5 vs CentOS7}
        \label{fig:pfs-c5vsc7-bias}
\end{subfigure}
\begin{subfigure}[b]{\columnwidth}
        \includegraphics[width=0.8\columnwidth]{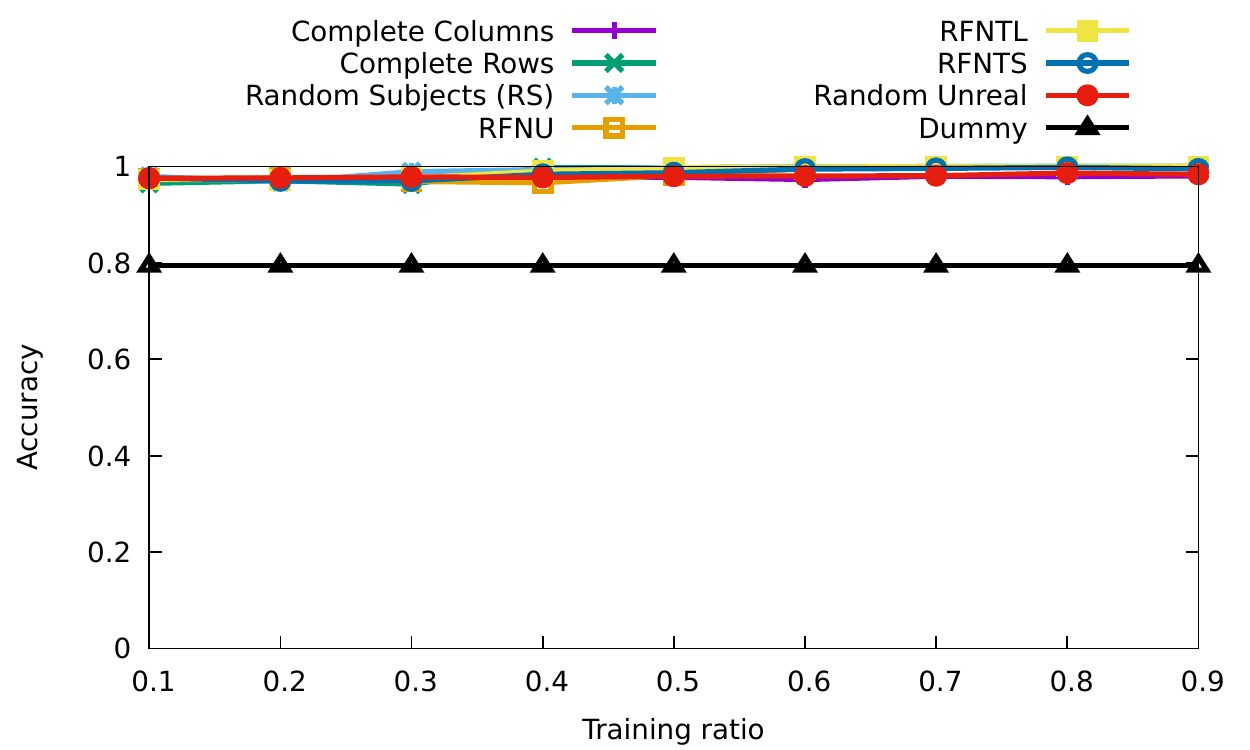}
        \caption{PFS, CentOS6 vs CentOS7}
\end{subfigure}\hfill
\begin{subfigure}[b]{\columnwidth}
        \includegraphics[width=0.8\columnwidth]{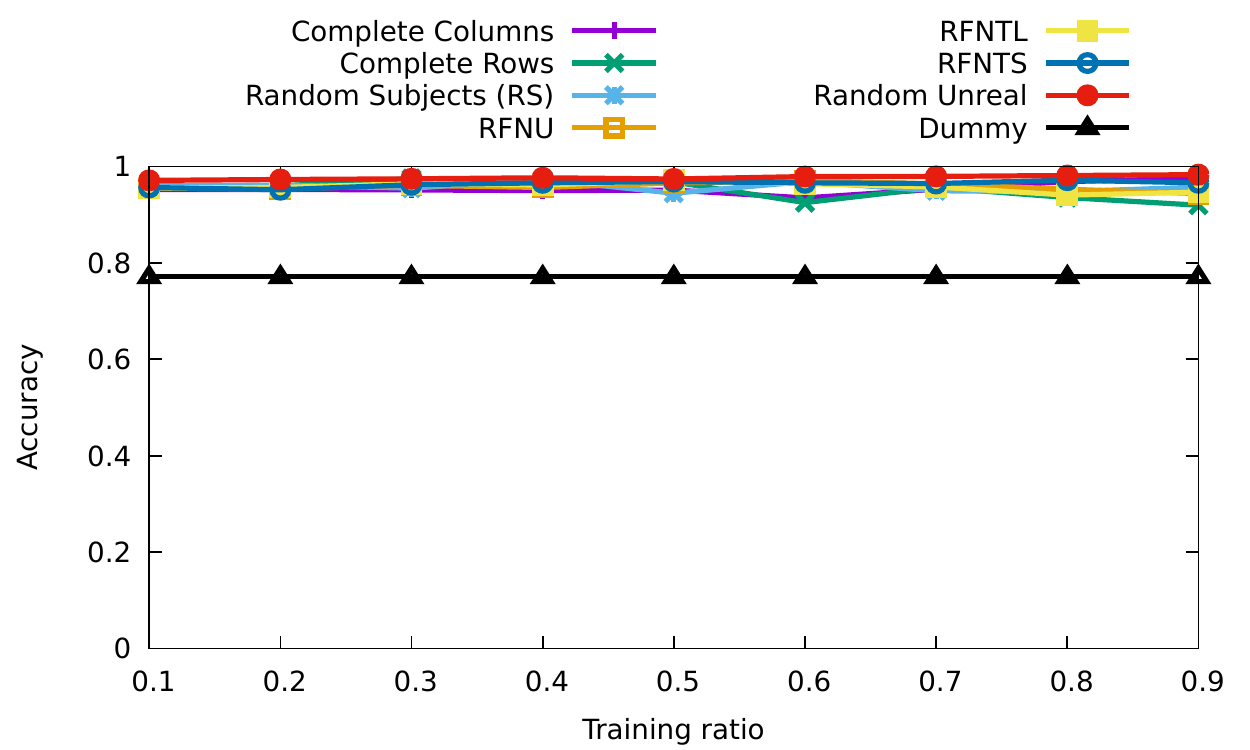}
        \caption{FS, CentOS6 vs CentOS7}
\end{subfigure}
\caption{Accuracy results on PreFresurfer (PFS) and Freesurfer (FS) data, ALS \emph{with} bias.}
\label{fig:results-real-als-bias}
\end{figure*}

Figure~\ref{fig:results-real-als-bias} shows the accuracy for ALS with 
bias. From a training ratio of 0.5, all methods perform well for all 
datasets, except complete columns in 
Figure~\ref{fig:pfs-c5vsc7-bias}. All methods except RFNU, RFNTL and 
RFNTS also perform well for a training ratio lower than 0.5. Overall, 
the accuracy is much higher than without bias, due to the fact that the 
file biases are very strong. In the remainder, all the 
results are obtained using ALS without bias.

\subsection{ROC analysis}

\begin{figure*}
\includegraphics[width=\columnwidth]{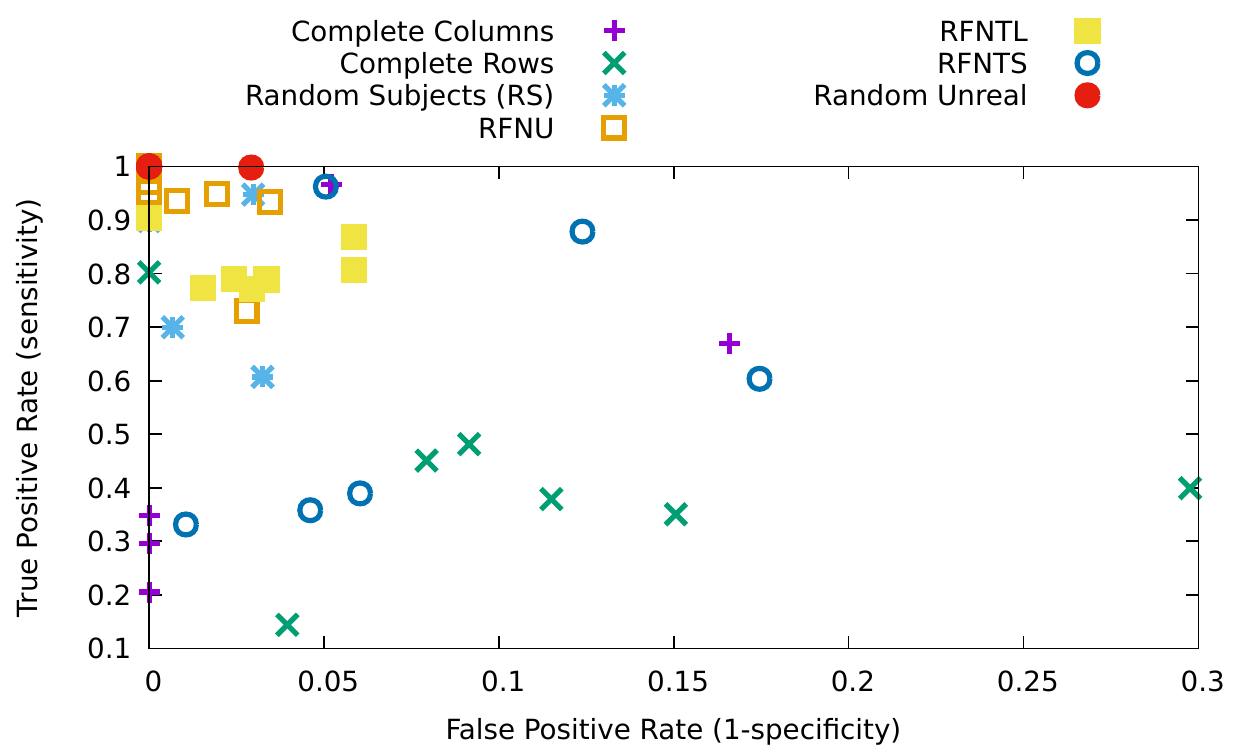}
\includegraphics[width=\columnwidth]{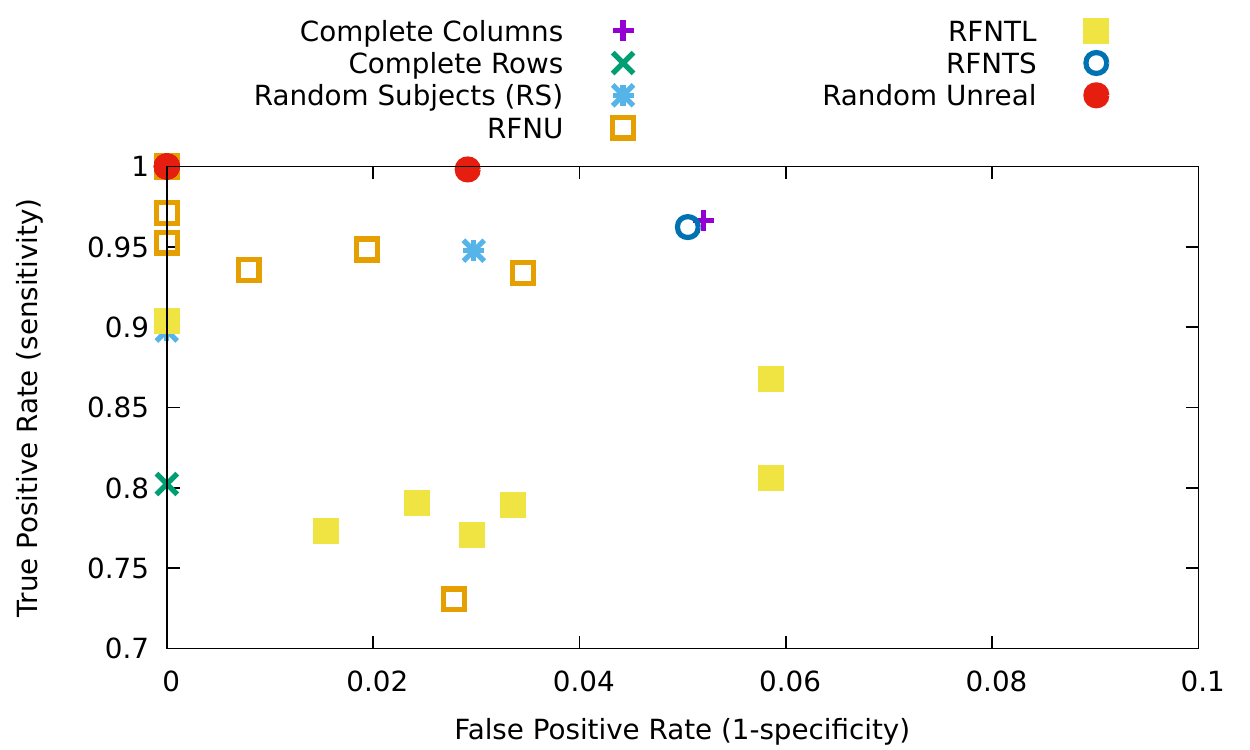}
\caption{Comparison of the sampling methods in the ROC space for the 6 synthetic datasets and Freesurfer (ALS without bias). Right: entire dataset. Left: close-up on the top-left part.}
\label{fig:roc}
\end{figure*}

Figure~\ref{fig:roc} compares the sampling methods in the ROC space, 
for a training ratio of 0.9, on the synthetic and Freesurfer dataset. 
The PreFreesurfer dataset is not included since specificity of RFNTL, 
RFNU, RS and complete rows is undefined at this training ratio (the 
test set only contains positive elements). The average sensitivity and specificity values 
are reported in Table~\ref{table:roc}, which confirms that RFNU is the best 
performing method on average.

\begin{table}
\centering
\begin{tabular}{ccc}
& Sensitivity & Specificity \\
\hline
Complete Columns & 0.53 & 0.97\\
Complete Rows & 0.43 & 0.89\\
Random Subjects & 0.88 & 0.99\\
RFNU & 0.92 & 0.99 \\
RFNTL & 0.81 & 0.97\\
RFNTS & 0.65 & 0.93 \\
Random Unreal & 1 & 1
\end{tabular}
\caption{Average sensitivity and specificity for the synthetic datasets and Freesurfer ($\alpha=0.9$, ALS without bias).}
\label{table:roc}
\end{table}

\subsection{Effect of the number of factors} 
\begin{figure*}
\begin{subfigure}{0.5\columnwidth}
\includegraphics[width=\columnwidth]{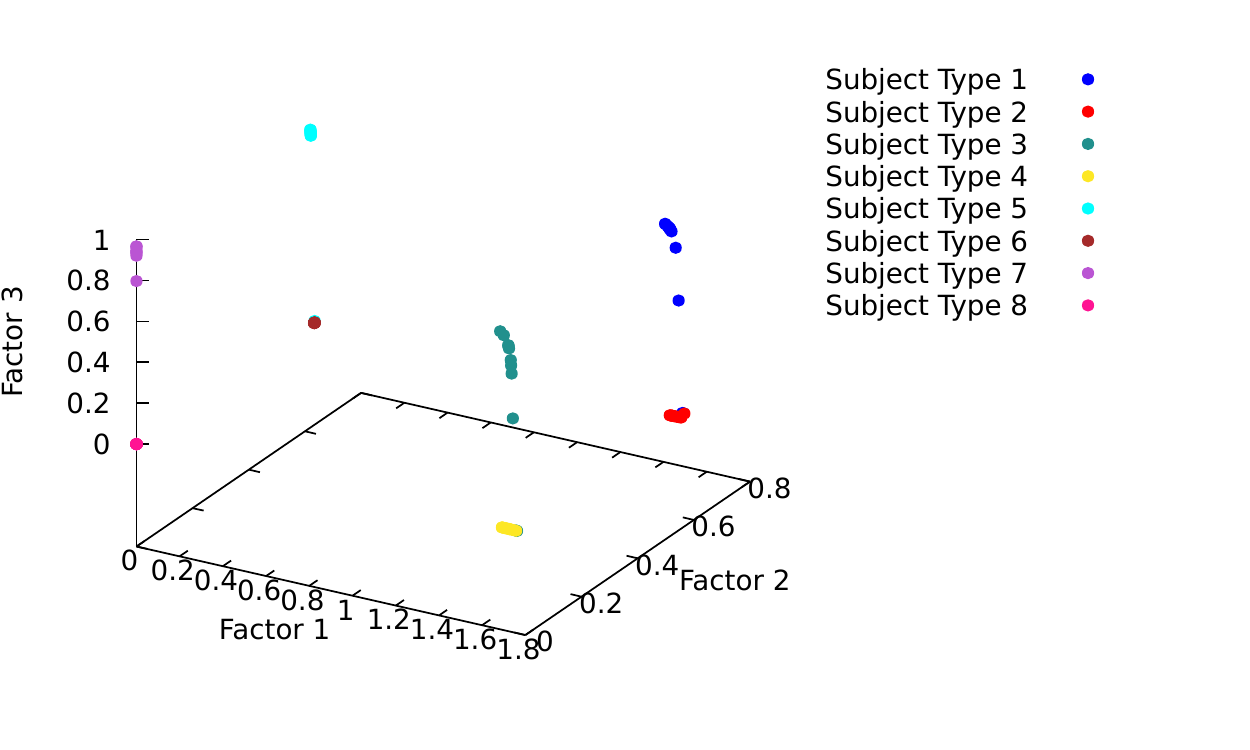}
\caption{Subject factors}
\end{subfigure}
\hfill
\begin{subfigure}{0.5\columnwidth}
\includegraphics[width=\columnwidth]{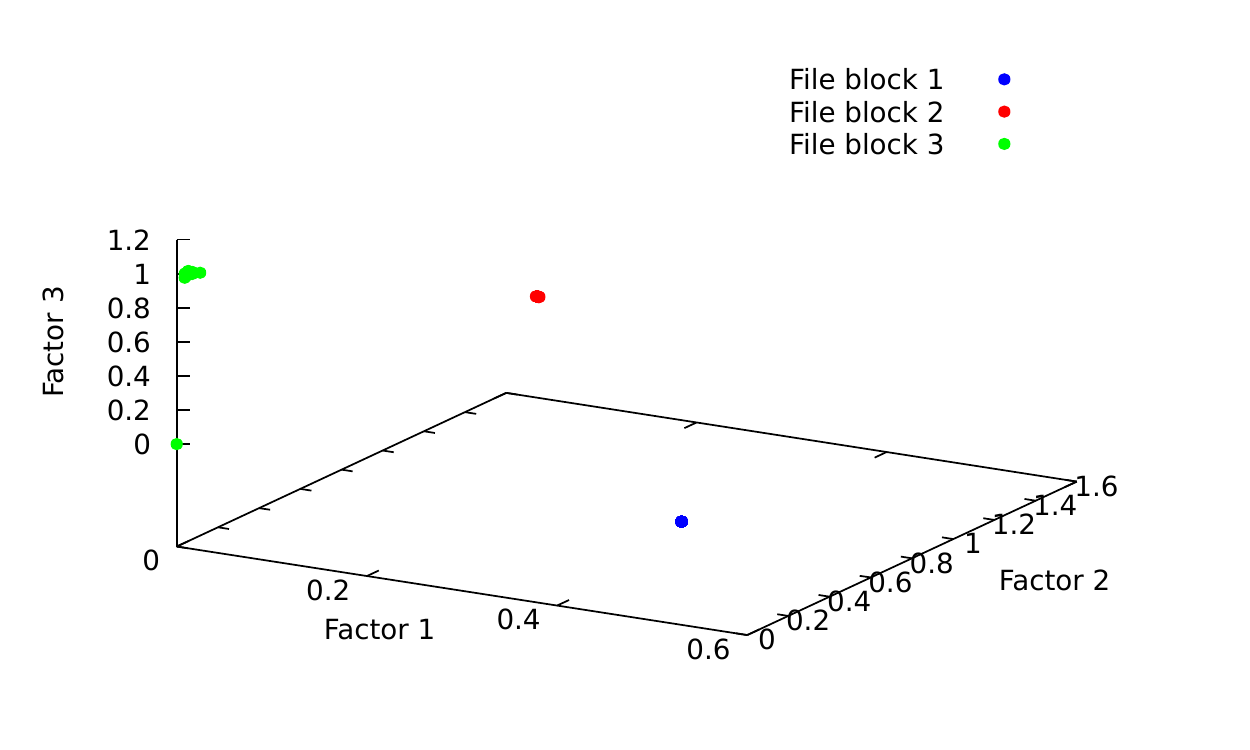}
\caption{File factors}
\end{subfigure}
\caption{Subject and file factors produced by collaborative filtering (RFNU, synthetic dataset, 8 types, 3 factors, 5 iterations).}
\label{fig:factors3d}
\end{figure*}

Figure~\ref{fig:factors} 
shows the effect of the number of factors used in the ALS optimization 
for the RFNU method and the synthetic dataset with 8 subject types (3 
blocks of files). For training ratios greater than 0.5, the accuracy 
with 3 factors is substantially higher than with 2 factors. Beyond 3, 
the number of factors does not have any effect, which shows that the 
data is best 
explained by 3 factors. Figure~\ref{fig:factors3d} confirms that 3 
factors are enough to separate the files of this dataset in 3 
blocks, and the subjects in 8 types.

\subsection{Effect of the maximum number of iterations}
\begin{figure}
\begin{minipage}{0.45\textwidth}
\includegraphics[width=\columnwidth]{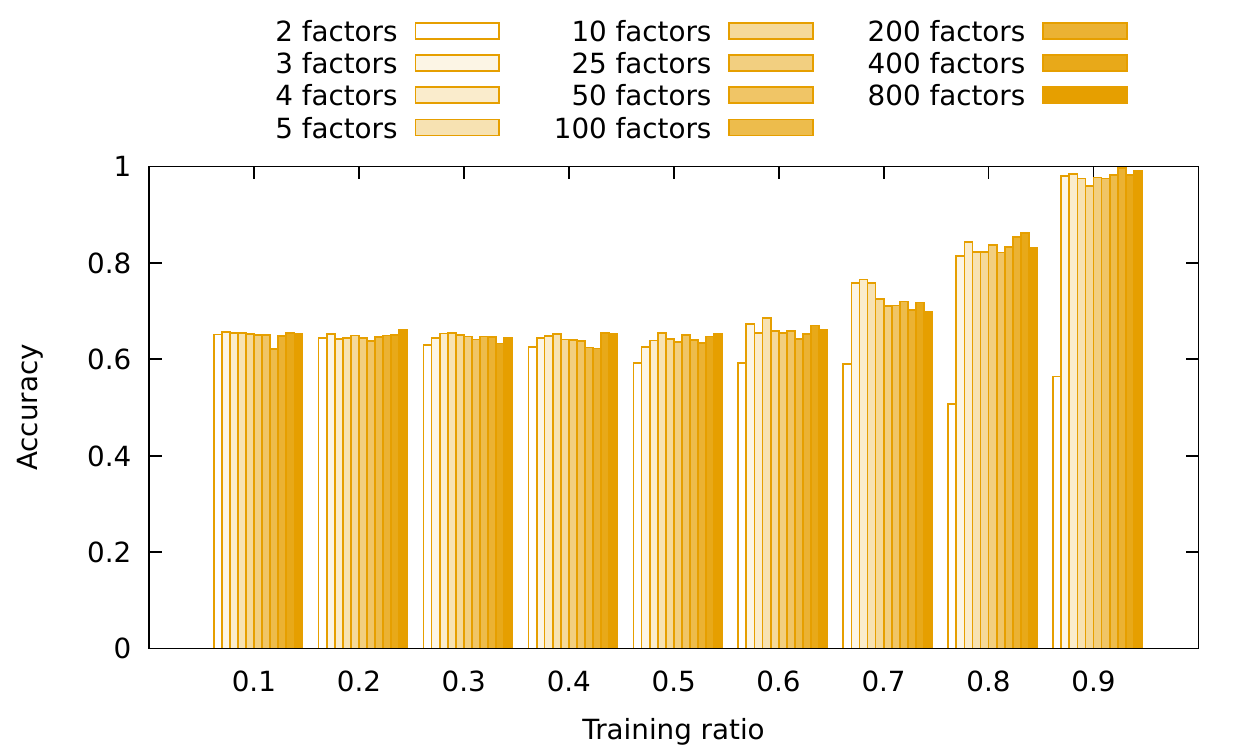}
\caption{Effect of the number of factors used in ALS, (RFNU, synthetic dataset, 8 types, 5 iterations).}
\label{fig:factors}
\end{minipage}\hfill
\begin{minipage}{0.45\textwidth}
\includegraphics[width=\columnwidth]{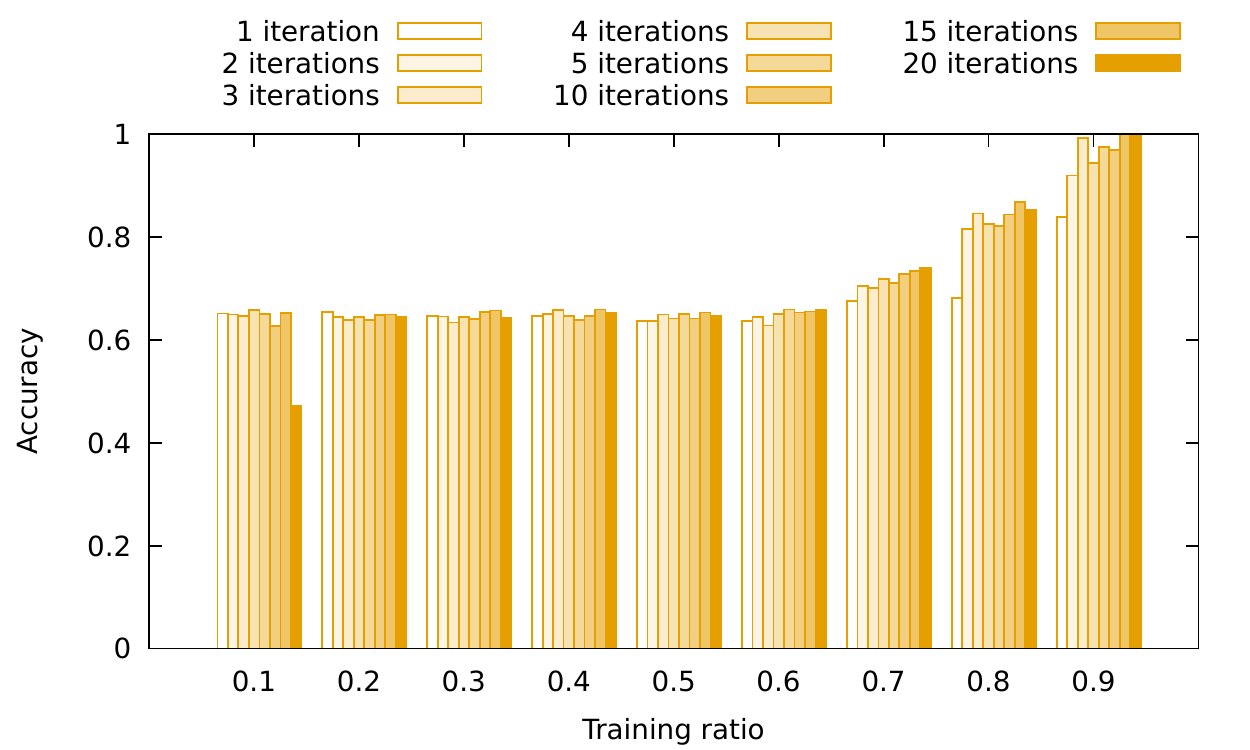}
\caption{Effect of the number of iterations used in ALS, (RFNU, synthetic dataset, 8 types, 50 factors).}
\label{fig:iterations}
\end{minipage}
\end{figure}

Figure~\ref{fig:iterations} illustrates the effect of the number of 
iterations used in the ALS optimization, for the 
RFNU method and the synthetic dataset with 8 types. 
From a training ratio of 0.7, the number of iterations has a moderate 
effect on the accuracy. We used 5 iterations in our experiments, 
which only slightly degrades the accuracy compared to 15 or 20.

\subsection{Prediction error localization}

\begin{figure*}
\begin{subfigure}[b]{0.3\columnwidth}
        \includegraphics[width=\columnwidth]{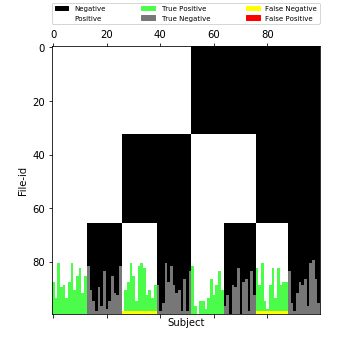}
        \caption{RFNU: $U(0.8N_f, N_f)$}
        \label{fig:overlap-rfnu}
\end{subfigure}
\hfill
\begin{subfigure}[b]{0.3\columnwidth}
        \includegraphics[width=\columnwidth]{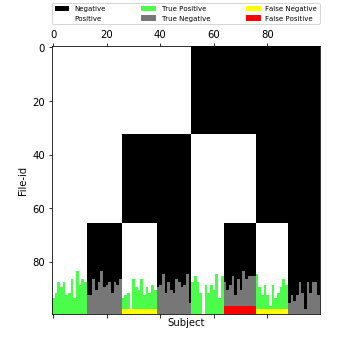}
        \caption{RFNTL: $T(0.85N_f, 0.85N_f, N_f)$}
        \label{fig:overlap-rfntl}
\end{subfigure}
\hfill
\begin{subfigure}[b]{0.3\columnwidth}
        \includegraphics[width=\columnwidth]{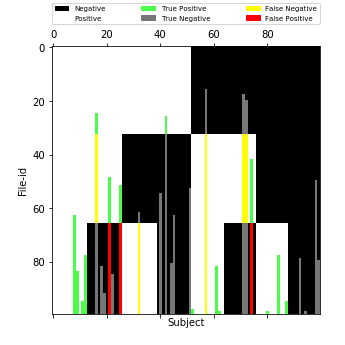}
        \caption{RFNTS: 0.3$T(0, N_f, N_f)$+0.7$N_f$}
        \label{fig:overlap-rfnts}
\end{subfigure}
\caption{Comparison between RFNU, RFNTL and RFNTS on synthetic dataset with 8 types, $\alpha=0.9$.}
\label{fig:overlap}
\end{figure*}

Figure~\ref{fig:overlap} compares the prediction errors made by RFNU, 
RFNTL and RFNTS. The training set is represented
in black (negatives) and white (positives), and the test set is in green 
(true positives), yellow (false negatives), gray (true negatives) and 
red (false positives). This representation provides insights regarding 
where, and perhaps why, prediction errors 
occur. At this training ratio, RFNU (Figure~\ref{fig:overlap-rfnu}) 
uniformly samples the number of files per subject between 0.8$N_f$ and 
$N_f$, which enables the training on the last 
files of the pipeline, while maintaining a low number of columns with 
a low training ratio. On the contrary, RFNTL 
(Figure~\ref{fig:overlap-rfntl}) samples the number of files per 
subject from 0.85$N_f$, but the probability to have a complete column 
is 0 (the one complete column in Figure~\ref{fig:overlap-rfntl} is the 
one included to prevent cold start issues), which leads to a ``stripe" 
of prediction errors at the bottom of the matrix. RFNTS 
(Figure~\ref{fig:overlap-rfnts}) does not have this issue, as it 
includes many complete columns in the training set. However, it comes at 
the cost of several columns with a number of files lower than 
60\%: in such columns, the prediction is often entirely wrong, which explains
the reduced accuracy compared to RFNU.

\begin{figure}[h]
\centering
\includegraphics[width=0.6\columnwidth]{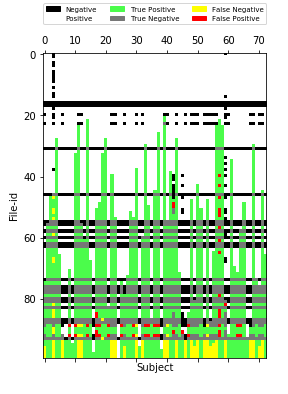}
\caption{Prediction results for RFNU on the Freesurfer dataset, $\alpha=0.6$.}
\label{fig:error-locality}
\end{figure}

Figure~\ref{fig:error-locality} shows the RFNU results on 
the Freesurfer dataset for a training ratio of 0.6. The prediction 
error is localized (1) at the bottom of the matrix, which corresponds 
to the end of the execution, and (2) in the regions where file 
reproducibility varies across subjects, i.e., lines are not entirely 
black or entirely white. This is consistent with our expectations and 
confirms the validity of our results.

\section{Conclusion}

While collaborative filtering, perhaps unsurprisingly, correctly 
predicts the missing values in a matrix modeling reproducibility 
evaluations, the usual random sampling method cannot be used in 
time-constrained processes. We proposed 6 sampling methods to address 
this issue, and we found that one of them, RFNU, performs 
better than the other ones on average. We 
explain that by the fact that RFNU builds the training set using a 
balanced mix of nearly complete and nearly empty columns, with a 
continuum of 
intermediate configurations. On the contrary, other 
methods, including RFNTL and RFNTS, bias the sampling toward complete 
or empty columns, which is sometimes detrimental to accuracy.

For datasets that are strongly dependent on row bias, such as the 
PreFreesurfer and Freesurfer ones, RFNU provides 
accuracy values consistently higher than 0.95 for training 
ratios higher than 0.5, even 
when biases are not included. For the synthetic dataset, RFNU still performs very well 
when biases are not included, but it does very poorly when biases are included.
For this reason, we recommend to not include biases in the collaborative filtering 
optimization to solve this problem. Even though biases provide a slight 
accuracy improvement when datasets are strongly biased (constant lines 
in the matrix), they can also be very detrimental for more complex 
datasets such as the synthetic one used here.

From a practical standpoint, our study shows that reproducibility 
evaluations of the PreFreesurfer and Freesurfer pipelines can be 
conducted using only 50\% of the files produced, with an accuracy above 
95\%. Potentially, this could reduce the computing time and storage 
required for such studies by a factor of 2. However, such studies could 
not be conducted by processing only half of the subjects entirely, 
which would correspond to the complete columns sampling method. 
Instead, the processing of all the subjects should be initiated and 
terminated in a random uniform way, assuming that files are uniformly 
produced throughout the execution. On more complex matrices, such as 
the synthetic dataset studied here, the training ratio required 
to get a 95\% accuracy increases to 0.85.

Our study could be extended to real-valued reproducibility 
matrices instead of just binary ones. It is indeed common for file 
differences to be quantified using specific similarity measures or 
distances, such as the Levenshtein distance between strings, or the sum of 
squared distances among voxels of an image. Our sampling methods 
could be directly applied to real-valued matrices, and we expect our 
conclusions on the best-performing sampling method (RFNU) and 
the inclusion of biases in the model to remain valid.


Finally, the method described in this paper could possibly be used to predict 
the outcome of other time-constrained processes, for instance 
markers of chronic disease activity.

\section*{Acknowledgement}

We thank Compute Canada for providing the compute and storage resources
required for this study.

\bibliographystyle{IEEEtran}
\bibliography{IEEEabrv,biblio}

\begin{thebibliography}{10}
\providecommand{\url}[1]{#1}
\csname url@samestyle\endcsname
\providecommand{\newblock}{\relax}
\providecommand{\bibinfo}[2]{#2}
\providecommand{\BIBentrySTDinterwordspacing}{\spaceskip=0pt\relax}
\providecommand{\BIBentryALTinterwordstretchfactor}{4}
\providecommand{\BIBentryALTinterwordspacing}{\spaceskip=\fontdimen2\font plus
\BIBentryALTinterwordstretchfactor\fontdimen3\font minus
  \fontdimen4\font\relax}
\providecommand{\BIBforeignlanguage}[2]{{%
\expandafter\ifx\csname l@#1\endcsname\relax
\typeout{** WARNING: IEEEtran.bst: No hyphenation pattern has been}%
\typeout{** loaded for the language `#1'. Using the pattern for}%
\typeout{** the default language instead.}%
\else
\language=\csname l@#1\endcsname
\fi
#2}}
\providecommand{\BIBdecl}{\relax}
\BIBdecl

\bibitem{peng2011reproducible}
R.~D. Peng, ``Reproducible research in computational science,'' \emph{Science},
  vol. 334, no. 6060, pp. 1226--1227, 2011.

\bibitem{baker2016there}
M.~Baker, ``Is there a reproducibility crisis? {A} {N}ature survey lifts the
  lid on how researchers view the crisis rocking science and what they think
  will help,'' \emph{Nature}, vol. 533, no. 7604, pp. 452--455, 2016.

\bibitem{gronenschild2012effects}
E.~H. Gronenschild, P.~Habets, H.~I. Jacobs, R.~Mengelers, N.~Rozendaal,
  J.~Van~Os, and M.~Marcelis, ``The effects of {FreeSurfer} version,
  workstation type, and macintosh operating system version on anatomical volume
  and cortical thickness measurements,'' \emph{PloS one}, vol.~7, no.~6, p.
  e38234, 2012.

\bibitem{glatard2015reproducibility}
T.~Glatard, L.~B. Lewis, R.~Ferreira~da Silva, R.~Adalat, N.~Beck, C.~Lepage,
  P.~Rioux, M.-E. Rousseau, T.~Sherif, E.~Deelman \emph{et~al.},
  ``Reproducibility of neuroimaging analyses across operating systems,''
  \emph{Frontiers in neuroinformatics}, vol.~9, p.~12, 2015.

\bibitem{van2013wu}
D.~C. Van~Essen, S.~M. Smith, D.~M. Barch, T.~E. Behrens, E.~Yacoub,
  K.~Ugurbil, W.-M.~H. Consortium \emph{et~al.}, ``The {WU-Minn} {Human
  Connectome Project}: an overview,'' \emph{Neuroimage}, vol.~80, pp. 62--79,
  2013.

\bibitem{bowring2018exploring}
A.~Bowring, C.~Maumet, and T.~Nichols, ``Exploring the impact of analysis
  software on task {fMRI} results,'' \emph{bioRxiv}, p. 285585, 2018.

\bibitem{feng2013efficient}
D.~Feng, C.~Germain, and T.~Glatard, ``Efficient distributed monitoring with
  active collaborative prediction,'' \emph{Future Generation Computer Systems},
  vol.~29, no.~8, pp. 2272--2283, 2013.

\bibitem{leskovec2014mining}
J.~Leskovec, A.~Rajaraman, and J.~D. Ullman, \emph{Mining of massive
  datasets}.\hskip 1em plus 0.5em minus 0.4em\relax Cambridge university press,
  2014.

\bibitem{breese1998empirical}
J.~S. Breese, D.~Heckerman, and C.~Kadie, ``Empirical analysis of predictive
  algorithms for collaborative filtering,'' in \emph{Proceedings of the
  Fourteenth conference on Uncertainty in artificial intelligence}.\hskip 1em
  plus 0.5em minus 0.4em\relax Morgan Kaufmann Publishers Inc., 1998, pp.
  43--52.

\bibitem{linden2003amazon}
G.~Linden, B.~Smith, and J.~York, ``Amazon.com recommendations: Item-to-item
  collaborative filtering,'' \emph{IEEE Internet computing}, no.~1, pp. 76--80,
  2003.

\bibitem{koren2009matrix}
Y.~Koren, R.~Bell, and C.~Volinsky, ``Matrix factorization techniques for
  recommender systems,'' \emph{Computer}, no.~8, pp. 30--37, 2009.

\bibitem{glasser2013minimal}
M.~F. Glasser, S.~N. Sotiropoulos, J.~A. Wilson, T.~S. Coalson, B.~Fischl,
  J.~L. Andersson, J.~Xu, S.~Jbabdi, M.~Webster, J.~R. Polimeni \emph{et~al.},
  ``The minimal preprocessing pipelines for the {Human Connectome Project},''
  \emph{Neuroimage}, vol.~80, pp. 105--124, 2013.

\end{thebibliography}

\newpage

\onecolumn

\end{document}